\documentclass[10pt]{article}
\usepackage{float}
\usepackage{amsmath}
\usepackage{mathtools}
\usepackage{amsfonts}
\usepackage{jheppub}
\usepackage{color}
\usepackage{subfigure}
\usepackage[utf8x]{inputenc}

\relax
\newcommand{\beq}{\begin{equation}}
\newcommand{\eeq}{\end{equation}}
\newcommand{\beqn}{\begin{eqnarray}}
\newcommand{\eeqn}{\end{eqnarray}}
\newcommand\be{\begin{equation}}
\newcommand\ee{\end{equation}}
\newcommand\bea{\begin{eqnarray}}
\newcommand\eea{\end{eqnarray}}
\def\mh{m_\Higgs}
\def\Higgs{h}
\def\tr{{\rm tr}}

\def\cM{{\cal M}}

\def\cA{{\cal A}}
\def\qb{\bar{q}}
\newcommand{\x}{\times}
\def\mC{\mathcal{C}}
\def\symbrack{}
\def\detS{\left|S_{1\x2\x3\x4}\right|}
\def\DeltaFour{G}
\def\spa#1.#2{\left\langle#1\,#2\right\rangle}
\def\s#1.#2{s_{#1#2}}
\def\spb#1.#2{\left[#1\,#2\right]}
\def\spab#1.#2.#3{\left\langle#1|#2|#3\right]}
\def\spba#1.#2.#3{\left[#1|#2|#3\right\rangle}
\def\spaa#1.#2.#3.#4{\left\langle#1|#2|#3|#4\right\rangle}
\def\spbb#1.#2.#3.#4{\left[#1|#2|#3|#4\right]}
\def\spaabb#1.#2.#3.#4.#5{\left\langle#1|#2|#3|#4|#5\right]}
\def\TrfourgL#1.#2.#3.#4{{\rm tr}_{-}\{{#1}\,{#2}\,{#3}\,{#4}\}}
\def\TrfourgR#1.#2.#3.#4{{\rm tr}_{+}\{{#1}\,{#2}\,{#3}\,{#4}\}}
\def\Trfour#1.#2.#3.#4{{\rm tr}\{{#1}\,{#2}\,{#3}\,{#4}\}}

\def\etildehat{\tilde{e}}
\def\DeltaThree{\Delta_3}

\author[a]{Lucy Budge,}
\emailAdd{lucy.budge@durham.ac.uk}

\author[b]{John M. Campbell,}
\emailAdd{johnmc@fnal.gov}

\author[a]{R. Keith Ellis,}
\emailAdd{keith.ellis@durham.ac.uk}

\author[c]{Satyajit Seth}
\emailAdd{seth@prl.res.in}

\affiliation[a]{Institute for Particle Physics Phenomenology, Durham University, Durham, DH1 3LE, UK}
\affiliation[b]{Fermilab, PO Box 500, Batavia IL 60510-5011, USA}
\affiliation[c]{Physical Research Laboratory, Navrangpura, Ahmedabad - 380009, India}

\date{\today}

\preprint{FERMILAB-PUB-20-476-T,\, IPPP/20/40}

\title{Analytic results for scalar-mediated Higgs boson production in association with two jets}

\abstract{We present compact analytic formulae for all one-loop amplitudes representing the production
  of a Higgs boson in association with two jets, mediated by a colour triplet scalar particle.
  Many of the integral coefficients present for scalar mediators
  are identical to the case when a massive fermion circulates in the 
  loop, reflecting a close relationship between the two theories.  
  The calculation is used to study Higgs boson production in association with two jets
  in a simplified supersymmetry (SUSY) scenario in which the dominant additional contributions
  arise from loops of top squarks.    
  The results presented here facilitate an indirect search for top squarks
  in this channel, by a precision measurement of the corresponding cross-section.
  However, we find that the potential for improved discrimination between
  the SM and SUSY cases suggested by the pattern of results in the 1- and 2-jet samples
  is unlikely to be realized due to the loss in statistical power compared to an inclusive analysis.}

\begin{document}
\maketitle

\section{Introduction}

The main production channel of Higgs bosons at the Large Hadron Collider (LHC)
is through gluon-gluon fusion {\em i.e.,} $gg\to h$.  The leading order process
starts at one loop, mediated by massive quark(s) of the Standard Model (SM).
By marrying high-quality data from run 2 of the LHC with precision
theoretical calculations for this
process~\cite{Anastasiou:2016cez,Mistlberger:2018etf}, one can extract
ever more exquisite determinations of the properties of the Higgs
boson~\cite{Sirunyan:2018kta,ATLAS:2020pvn}.

As more data is collected, additional information can be obtained from analyzing
differential information beyond inclusive cross sections.  One reason this is
important is that additional jet activity allows new kinematic regions to be
examined that may be more sensitive probes of Higgs properties.  An example
of this is that the nature of the Higgs coupling to particles circulating
in the loop can only be probed if the relevant energy scale is at least of
order of the particle's mass.  For inclusive production the relevant energy scale
is the Higgs mass and, since $\mh < m_t$, one can describe this process
using an effective field theory (EFT) in which the loop of heavy top quarks
is replaced by an effective Lagrangian,
\begin{equation}
\label{EFT}
{\mathcal L}_{{\rm eff}} = \frac{g_s^2}{48\pi^2 v} \, \Higgs \, G_{\mu\nu}^A G^{A,\mu\nu} \, ,
\end{equation}
where $g_s$ is the strong coupling constant, $v$ is the vacuum expectation value of the Higgs field,
$G_{\mu\nu}$ is the QCD field strength, and $\Higgs$ is the Higgs boson field.
Indeed, the efficacy of this approximation is the very reason that such
high-precision calculations of this process can be
performed~\cite{Anastasiou:2016cez,Mistlberger:2018etf}.
In the presence of additional jet activity the relevant energy scale is no
longer $\mh$ but is instead the transverse momentum ($p_T$) of the leading jet.
Therefore the Higgs+jet process can become especially sensitive to the coupling of
the Higgs boson to new mediator particles of mass $m_X$ once $p_T > m_X$.
An analysis of the cross section for this process, in this kinematic regime, could
thus provide the first signal of new physics (in the case of a deviation from the SM prediction),
or a stringent bound on the mass and coupling of any new mediator particle.
Although less sensitive than corresponding direct searches, such indirect probes
of the mediator particles are insensitive to any assumptions regarding the nature
of their decay chains and may therefore provide complementary information.

Massive colour triplet scalar particles that arise in beyond the
Standard Model (BSM) scenarios, are potential new mediators for
couplings of gluons to the Higgs boson. Indeed, one of the main goals
of the LHC is to further explore the particles to which the Higgs
boson couples, and having already discovered one fundamental scalar
particle it is natural to consider whether further scalar degrees of
freedom might exist.  One such proposed scalar particle is the top
squark, a super partner of the SM top quark that appears in the
Minimal Supersymmetric Standard Model (MSSM).  The effect of such
loops of particles has been explored
previously~\cite{Bonciani:2007ex,Brein:2007da,Grojean:2013nya},
focussing on effects in either inclusive Higgs production or in the
case of the Higgs boson recoiling against a single jet.  Most
recently, Ref.~\cite{Banfi:2018pki} demonstrated that the 1-jet
process offers, in principle, superior information to inclusive
production over certain regions of parameter space.  For the top
squark, current indirect limits from Higgs and electroweak
data~\cite{Espinosa:2012in,Ellis:2018gqa} are around $m_{\tilde t}
\sim 300$~GeV.  As noted above, this limit is clearly much weaker than
any direct limit derived from a specific decay chain, which is
currently around the 1~TeV scale (see, for example,
Refs.~\cite{Sirunyan:2020tyy} and~\cite{Aaboud:2017aeu} for recent
limits from CMS and ATLAS).

In this paper we will extend this analysis to the case in which a
Higgs boson is produced in association with two jets.  To do so we
have performed a new calculation of the amplitudes for the scattering
of a Higgs boson with four partons, mediated by a loop of coloured
scalar particles.  Our results are expressed in the form of compact
analytic expressions, exploiting a close correspondence with their
fermionic counterparts~\cite{Budge:2020oyl}.  The resulting
expressions may be evaluated numerically in a fast and stable manner,
allowing for the construction of an efficient Monte Carlo event
generator.  We first outline the generic scalar theory in which we
shall perform our calculation, as well as the specific MSSM case, in
section~\ref{sec:setup}.  The computation of the four-parton matrix
elements entering the Higgs+2~jet analysis is given in
section~\ref{sec:calculation}, with results for one-loop integral
coefficients for the case of a scalar loop detailed in
section~\ref{sec:coeffs}.  We move to phenomenology in
section~\ref{sec:recap}, first providing a recalculation and recap of
results for the 0- and 1-jet cases before presenting our new 2-jet
analysis in section~\ref{sec:pheno}.  Our conclusions are drawn in
section~\ref{sec:conclusions}.  Finally, as an aid to performing an
independent implementation of the formulae presented here,
appendix~\ref{sec:PScheck} provides numerical results for the integral
coefficients given in section~\ref{sec:coeffs}, and
appendix~\ref{sec:largemass} details the connection between our
amplitudes and those obtained in the EFT.

\section{Setup}
\label{sec:setup}

We first formulate a generic scalar theory involving a complex scalar $\phi$ which
carries $SU(3)$ colour in the triplet representation.  The Lagrangian involving
$\phi$ thus reads, 
\begin{equation} \label{ScalarLagrangian}
{\cal L} = (D^\mu \phi_i^\dagger) (D_\mu \phi_i) -\lambda \phi^\dagger_i \phi_i \Higgs,\;\;\;\;
D_\mu \phi_i=\partial_\mu\phi_i + i \frac{g_s}{\sqrt{2}} (t \cdot {\cal G}_\mu)_{ij} \phi_j\,,
\end{equation}
where 
${\cal G_{\mu}}^{\!\!\!a}$ denotes the
gluon field, $t^a$ represents the standard $SU(3)$ colour generators, normalized 
such that $\tr(t^a t^b)\;=\; \delta^{ab}$ and $g_s$ is the strong coupling constant. The
coupling of the Higgs boson to the scalar field is denoted by the parameter $\lambda$.

\subsection{Overview}

In order to elucidate the differences -- and similarities -- between the cases of
Higgs boson production mediated by a fermion and a scalar loop, we first examine
the amplitudes for inclusive Higgs boson production via these two processes. 

In the Standard Model, where the particle in the loop is a quark of mass $m$, the amplitude
for $g(p_1) g(p_2) \to h$ takes the following form,
\beqn  \label{Fermion2gluonresult}
{\cal H}^{gg}_2 &=& i\, \frac{g_s^2}{16\,\pi^2}\, \delta^{AB}\, \Big(\frac{m^2}{v}\Big)
   \Big[g^{\mu\nu}-\frac{p_1^{\nu}\,p_2^{\mu}}{p_1.p_2}\Big]
   \Big[(2\, \mh^2- 8 m^2)\, C_0(p_1,p_2;m)-4\Big] 
   \epsilon_\mu(p_1) \epsilon_\nu(p_2) \, .
\eeqn
The gluons have colour labels $A$ and $B$, $\epsilon$ represents a polarization vector, 
and $\mh$ denotes the mass of the Higgs boson.  The integral over the loop momentum is
encapsulated in the scalar triangle function $C_0(p_1, p_2, m)$ defined later in Eq.~(\ref{Integral_defns}).
The corresponding result for a loop containing a scalar
particle of mass $m$ is~\cite{Dawson:1996xz},
\beqn  \label{Scalar2gluonresult1}
{\cal A}^{gg}_2 &=& i\, \frac{g_s^2}{16\,\pi^2}\, \delta^{AB}\,\Big(\frac{-\lambda}{4}\Big) \Big[g^{\mu\nu}-\frac{p_1^{\nu}\,p_2^{\mu}}{p_1.p_2}\Big]
       \Big[ -8 m^2 \, C_0(p_1,p_2;m)-4\Big]  
   \epsilon_\mu(p_1) \epsilon_\nu(p_2)\, .
\eeqn
Writing the amplitudes in this way highlights several similarities between them.  When
setting $(-\lambda/4) = m^2/v$ the coefficient of the triangle integral proportional to $m^2$
is identical, as well as the purely rational term ($-4$).  This illustrates
a general correspondence between such coefficients in the scalar and fermion
theories~\cite{Budge:2020oyl}.

\begin{figure}[t]
\begin{center}
\includegraphics[angle=270,width=0.7\textwidth]{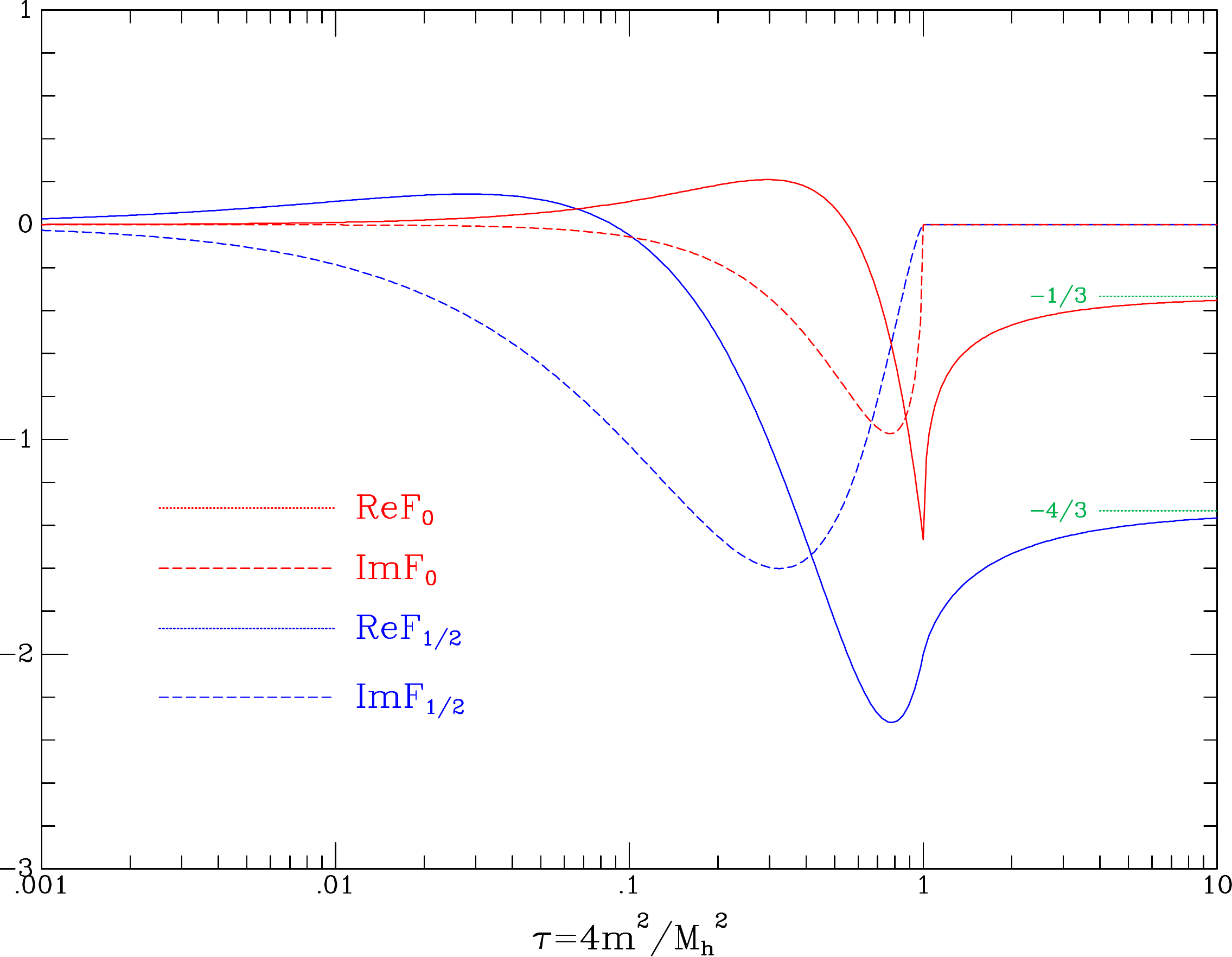}
\caption{The functions $F_{1/2}$ and $F_0$, given in Eqs.~(\ref{Fhalf})
and~(\ref{Fzero}) respectively,  plotted as a function of their arguments.
The (green) dashed lines show their asymptotes at large $\tau$. }
\label{figure:F}
\end{center}
\end{figure}

It is instructive to push this comparison further by extracting an overall factor as follows,
\beqn  \label{Scalar2gluonresult2}
{\cal H}^{gg}_2 &=& i\, \frac{g_s^2}{16\,\pi^2}\, \delta^{AB}\, 
\Big[\frac{1}{2}\big(g^{\mu\nu}-\frac{p_1^{\nu}\,p_2^{\mu}}{p_1.p_2}\big)\Big]
 \, \Big(\frac{\mh^2}{v}\Big) \epsilon_\mu(p_1) \epsilon_\nu(p_2) \, F_{1/2}(\tau) \, , \nonumber \\
{\cal A}^{gg}_2 &=& i\, \frac{g_s^2}{16\,\pi^2}\, \delta^{AB}\,
\Big[\frac{1}{2}\big(g^{\mu\nu}-\frac{p_1^{\nu}\,p_2^{\mu}}{p_1.p_2}\big)\Big]
 \,\Big(\frac{\lambda}{2}\frac{\mh^2}{m^2}\Big) \epsilon_\mu(p_1) \epsilon_\nu(p_2) \, F_{0}(\tau) \, ,
\eeqn
where the functions for the scalar and fermionic cases are given by,
\begin{eqnarray}
F_{0}(\tau) &=&  \tau \big[ 1- \tau f(\tau)\big] \label{Fzero} \, , \\
F_{1/2}(\tau) &=& -2 \tau  \big[1+(1-\tau)f(\tau)\big] \label{Fhalf},\;\;\; F_{1/2}(\tau)=-2 F_{0}(\tau)-2\tau f(\tau)\,,
\end{eqnarray}
and $\tau=4 m^2/\mh^2$.  In these formulae we have introduced the triangle
function $f(\tau) = -\frac{\mh^2}{2} \, C_0(p_{1},p_{2};m)$, for which the explicit result is, 
\beq
f(\tau) = -\frac{1}{4}\theta(1-\tau) \left[ \ln\left(
{1+\sqrt{1-\tau}\over 1-\sqrt{1-\tau}}\right) -i\pi\right]^2  
     + \theta(\tau-1)\left[\sin^{-1}(1/\sqrt{\tau})\right]^2 \; .
\label{fx}
\eeq
Figure~\ref{figure:F} shows the behaviour of $F_0$ and $F_{1/2}$ as a function of $\tau$.
In the region $\tau > 1$ the functions are both real and both negative, and quickly
approach their asymptotic values of $-1/3$ (scalar) and $-4/3$ (fermion). For the SM case
$\tau \gg 1$ and the value of this function is in the asymptotic regime, motivating the
use of the EFT shown in Eq.~(\ref{EFT}).  From these asymptotic values and the overall
factors extracted in Eqs.(\ref{Fhalf}) and~(\ref{Fzero}) it is clear that the two processes
can be described by the same effective Lagrangian, in the limit of large intermediary mass,
when we have $(\lambda/8) = m^2/v$.

\subsection{Top squarks in the MSSM}
\label{MSSMsection}
Although the analytic results presented later in this paper are valid for a generic color-triplet scalar,
and may be applicable more generally, in this paper we will focus on the top squark sector of the MSSM,
which contains two such scalars, $\tilde t_1$ and $\tilde t_2$.
We will consider scenarios in which the coupling of the lightest
Higgs boson in the MSSM is modified, assuming that this corresponds to the particle already observed
at the LHC.  Each squark couples to the lightest Higgs boson through a contribution
to the Lagrangian of the form shown in Eq.~(\ref{ScalarLagrangian}), where we now label the strength
of the Higgs coupling to each scalar by $\lambda_{h \tilde t_1 \tilde t_1}$ and
$\lambda_{h \tilde t_2 \tilde t_2}$ respectively.   Following ref.~\cite{Banfi:2018pki} we parametrize
this sector by,
\bea
(m_{\tilde{t}_1}  \ , \,  \Delta m \ , \,  \theta),\,\,\,\,\Delta m=\sqrt{m^2_{\tilde{t}_2}-m^2_{\tilde{t}_1}} \, ,
\eea
where, rather than using the two squark masses, we use the lightest top squark mass ($m_{\tilde{t}_1}$)
and a measure of the separation with the other state ($\Delta m$).  The final parameter ($\theta$) is
the mixing angle between the two scalar states, which takes values in the range $[-\pi/2,\pi/2]$.
If the mass of the MSSM pseudoscalar ($A$) is much larger than the weak scale ($m_A \gg m_Z$)
we can work in the decoupling limit~\cite{Gunion:2002zf}, in which the Higgs-squark couplings take
a very simple form:
\bea
\lambda_{h\, \tilde{t}_1 \tilde{t}_1} &=& \frac{m_t^2}{v} \, \left(
 \alpha_1 \cos^2\theta+ \alpha_2 \sin^2\theta +  2-\frac{(\Delta m)^2}{2 m_t^2} \sin^2 2\theta   \right)\,, \\
\lambda_{h\, \tilde{t}_2 \tilde{t}_2} &=& \frac{m_t^2}{v} \, \left(
 \alpha_1 \sin^2\theta+ \alpha_2 \cos^2\theta +   2+\frac{(\Delta m)^2}{2 m_t^2} \sin^2 2\theta   \right) \ .
\eea
The coefficients $\alpha_1$ and $\alpha_2$ in these formulae are given by,
\bea
\alpha_1 &=&  \frac{m_Z^2}{m_t^2} \cos2 \beta \left( 1-\frac{4}{3} \sin^2 \theta_W\right) \,, \\
\alpha_2 &=&  \frac{4}{3} \frac{m_Z^2}{m_t^2} \cos 2 \beta \sin^2\theta_W  \ ,
\eea
where $\theta_W$ is the weak mixing angle.  These formulae also contain the final MSSM parameter that is
necessary to specify our model, $\beta$, where $\tan\beta$ is the ratio of the vacuum expectation value
of the two Higgs bosons.

\section{Calculation of Higgs+2 jet process}
\label{sec:calculation}

We now provide the details of our calculation of the four-parton amplitudes
that enter our analysis of the Higgs+2 jet process.

\subsection{Amplitudes for a scalar loop}
We begin with the Lagrangian given in Eq. (\ref{ScalarLagrangian}) and break the amplitude 
for the production of a Higgs boson and $n$ gluons, mediated by a scalar loop,
into colour-ordered sub-amplitudes.  Following the notation of Ref.~\cite{Budge:2020oyl} we have,
\begin{eqnarray}
        \label{exp}
        \cA^{gggg}_n(\{p_i,h_i,c_i\})\,&=&\,i\frac{g_s^n}{16 \pi^2} \left(-\frac{\lambda}{4}\right)
	\sum_{\{1,2,\dots,n\}'}\;\tr\,(t^{c_1}t^{c_2} \dots t^{c_n})
        A_n^{\{c_i\}}(1^{h_1},2^{h_2},\ldots n^{h_n};\Higgs)\, ,
\end{eqnarray}
where the sum with the {\it prime}, $\sum_{\{1,2,\dots,n\}'}$, is over all
$(n-1)!$ {\em non-cyclic}  permutations of $1,2,\dots,n$, $\lambda$ is the
Higgs-scalar-scalar coupling and the $t$ matrices are the SU(3) matrices 
in the fundamental representation normalized such that,
\begin{equation}    \label{normalization}
        \tr(t^a t^b)\;=\; \delta^{ab}.
\end{equation}
$m$ is the mass of the scalar circulating in the loop.
It is sufficient to calculate one permutation in this sum, with the other colour sub-amplitudes
related by Bose symmetry and obtained by exchange.
The explicit result for the four gluon case is,
\begin{eqnarray}
  \label{explicitfor4}
  \cA^{gggg}_4(\{p_i,h_i,c_i\})\,&=&\,i\frac{g_s^4}{16 \pi^2}  \bigg(\frac{-\lambda}{4}\bigg)
\Bigg\{\big[ \tr\,(t^{c_1}t^{c_2}t^{c_3}t^{c_4})+ \tr\,(t^{c_1}t^{c_4}t^{c_3}t^{c_2}) \big] A_4^{1 2 3 4}(1^{h_1},2^{h_2},3^{h_3},4^{h_4};\Higgs) \nonumber \\
 & +&\big[ \tr\,(t^{c_1}t^{c_3}t^{c_4}t^{c_2})+ \tr\,(t^{c_1}t^{c_2}t^{c_4}t^{c_3}) \big] A_4^{1 3 4 2}(1^{h_1},2^{h_2},3^{h_3},4^{h_4};\Higgs) \nonumber \\
       &+&\big[ \tr\,(t^{c_1}t^{c_4}t^{c_2}t^{c_3})+ \tr\,(t^{c_1}t^{c_3}t^{c_2}t^{c_4}) \big] A_4^{1 4 2 3}(1^{h_1},2^{h_2},3^{h_3},4^{h_4};\Higgs) \Bigg\} \, .
\end{eqnarray}

We also need the amplitude for the production of a Higgs boson, an antiquark,
quark and two gluons.  It can be similarly decomposed into colour-ordered amplitudes as follows. 
\begin{eqnarray}
        \label{explicitfor2q}
{\cA}^{{\qb}qgg}_4(\{p_i,h_i,c_i,j_i\})= i\, \frac{g_s^4}{16 \pi^2 } \bigg(\frac{-\lambda}{4}\bigg) && \Big[
  (t^{c_3}\,t^{c_4})_{j_2\,j_1} A^{34}_4(1^{h_1},2^{-h_1},3^{h_3},4^{h_4};\Higgs) \nonumber \\
&&+(t^{c_4}\,t^{c_3})_{j_2\,j_1} A^{43}_4(1^{h_1},2^{-h_1},3^{h_3},4^{h_4};\Higgs) \Big] \;.
\end{eqnarray}
The colour structure $\delta^{c_3\,c_4}\,\delta_{j_2\,j_1}/N$ is also present in 
individual diagrams but makes no net contribution to the one-loop amplitude.
Here we will give results for the colour-ordered amplitude $A_4^{34}$ since it is straightforward
to obtain $A_4^{43}$ from this through the parity operation (complex conjugation) and permutation
of momentum labels.

The four-quark amplitude takes the form,
\begin{equation}
{\cA}^{4q}_4(\{p_i,h_i,j_i\})= i\, \frac{g_s^4}{16 \pi^2 } \bigg(\frac{-\lambda}{4}\bigg)\; (t^{c_1})_{j_2\,j_1}\;(t^{c_1})_{j_4\,j_3} 
A^{4q}_{4}(1^{h_1}_{{\qb}},2^{-h_1}_q,3^{h_3}_{{\qb}^\prime},4^{-h_3}_{q^\prime}) \, ,
\end{equation}
where the helicities of the quarks are fixed by those of the antiquarks.

All colour subamplitudes are then decomposed in terms of scalar integrals.  For instance, for the
Higgs + 4 gluon case we have,
\begin{eqnarray} \label{scalarreduction4gluons}
A_4^{1234}(1^{h_1},2^{h_2},3^{h_3},4^{h_4};\Higgs) & = & \frac{\bar\mu^{4-n}}{r_\Gamma}\frac{1}{i \pi^{n/2}} \int {\rm d}^n \ell
 \, \frac{{\rm Num}(\ell)}{\prod_i d_i(\ell)} \nonumber \\
&=& \sum_{i,j,k} \tilde{d}_{i\x j\x k}(1^{h_1},2^{h_2},3^{h_3},4^{h_4}) \, D_0(p_i, p_j, p_k ;m)  \nonumber \\
&+& \sum_{i,j} \tilde{c}_{i\x j}(1^{h_1},2^{h_2},3^{h_3},4^{h_4}) \,  C_0(p_i,p_j ;m)   \nonumber \\
&+& \sum_{i} \tilde{b}_{i}(1^{h_1},2^{h_2},3^{h_3},4^{h_4}) \, B_0(p_i;m) + \tilde{r}(1^{h_1},2^{h_2},3^{h_3},4^{h_4})\, .
\end{eqnarray}
$\tilde{r}$ are the rational terms and the sums in the above equations scan over groupings of
external gluons.  The scalar bubble ($B_0$), triangle ($C_0$) and box ($D_0$) integrals
are defined by,
\begin{eqnarray}
\label{Integral_defns}
B_0(p_1;m) &=& \frac{\bar\mu^{4-n}}{r_\Gamma}\frac{1}{i \pi^{n/2}} \int {\rm d}^n\ell \, \frac{1}{D(\ell) D(\ell_1)} \, ,\nonumber \\
C_0(p_1,p_2;m) &=& \frac{1}{i \pi^{2}} \int {\rm d}^4\ell \, \frac{1}{D(\ell) D(\ell_1) D(\ell_{12})} \, ,\nonumber \\
D_0(p_1,p_2,p_3;m) &=& \frac{1}{i \pi^{2}} \int {\rm d}^4\ell \, \frac{1}{D(\ell) D(\ell_1) D(\ell_{12})D(\ell_{123})} \, ,\nonumber \\
E_0(p_1,p_2,p_3,p_4;m) &=& \frac{1}{i \pi^{2}} \int {\rm d}^4\ell \, \frac{1}{D(\ell) D(\ell_1) D(\ell_{12}) D(\ell_{123}) 
D(\ell_{1234})} \, ,
\end{eqnarray}
where the denominators are defined as
\begin{equation} \label{denominatordef}
D(\ell) = \ell^2-m^2+i\varepsilon \, ,
\end{equation}
and the propagator momenta are,
\begin{eqnarray} \label{denominators}
\ell_1 &=& \ell+p_1 =\ell+q_1\, , \nonumber \\
\ell_{12} &=& \ell+p_1+p_2 =\ell+q_2\, , \nonumber \\
\ell_{123}&=& \ell+p_1+p_2+p_3 =\ell+q_3\, , \nonumber \\
\ell_{1234} &=& \ell+p_1+p_2+p_3+p_4 =\ell+q_4\, .
\end{eqnarray}
Finally, $r_\Gamma=1/\Gamma(1-\epsilon)+O(\epsilon^3)$ and $\bar\mu$ is an
arbitrary mass scale.

As explained in Ref.~\cite{Budge:2020oyl}, we have chosen to work in a basis without pentagon
integrals.  Nevertheless, we are left with some vestiges of their presence through pentagon-to-box
reduction coefficients, $\mC_{1\times2\times3\times4}^{(i)}$.  These can be written as,
\begin{eqnarray}
\label{pentredcoeffs}
\mC_{1\times2\times3\times4}^{(1)}
&=&-\frac{1}{2}\,\frac{s_{23}\,s_{34}\,[2\,s_{12}\,s_{24}+s_{13}\,s_{24}+s_{34}\,s_{12}-s_{23}\,
s_{14}]}{16\, \detS} \, , \nonumber\\
\mC_{1\times2\times3\times4}^{(2)}
&=&-\frac{1}{2}\,\frac{s_{34}\,[s_{1234}\,s_{23}\,(s_{123}-2\,s_{12})+s_{123}\,(s_{34}\,(s_{123}
-s_{23})+s_{12}\,(s_{234}+s_{23})-s_{234}\,s_{123})]}{16\, \detS} \, , \nonumber\\
\mC_{1\times2\times3\times4}^{(3)}
&=&-\frac{1}{2}\,\frac{[s_{14}\,s_{23}-(s_{12}+s_{13})\,(s_{24}+s_{34})]\,[s_{34}\,s_{12}+s_{23}
\,s_{14}-s_{13}\,s_{24}]}{16\, \detS} \, , \nonumber\\
\mC_{1\times2\times3\times4}^{(4)}
&=&-\frac{1}{2}\,\frac{s_{12}\,[s_{1234}\,s_{23}\,(s_{234}-2\,s_{34})+s_{234}\,(s_{12}\,(s_{234}-s_{23})+s_{34}\,(s_{123}+s_{23})-s_{234}\,s_{123})]}{16\, \detS} \, , \nonumber\\
\mC_{1\times2\times3\times4}^{(5)}
&=&-\frac{1}{2}\,\frac{s_{12}\,s_{23}\,[2\,s_{34}\,s_{13}+s_{13}\,s_{24}+s_{34}\,s_{12}-s_{23}\,s_{14}]}{16\, \detS} \, .
\end{eqnarray}
The denominator factor $\detS$ is the determinant of the matrix,
$\left[S_{1\x2\x3\x4}\right]_{ij}=[m^2-\frac{1}{2}(q_{i-1}-q_{j-1})^2]$,
where $q_i$ is the offset momentum, see Eq.~(\ref{denominators}).  It is given by,
\begin{eqnarray}
\label{Sdef}
16\, \detS&=&s_{12}\,s_{23}\,s_{34}\,\big(s_{14}\,s_{23}-(s_{12}+s_{13})\,(s_{24}+s_{34})\big)+m^2\,\DeltaFour\,,\nonumber \\
\DeltaFour  &=& (s_{12}\,s_{34}-s_{13}\,s_{24}-s_{14}\,s_{23})^2-4\,s_{13}\,s_{14}\,s_{23}\,s_{24} \,.
\end{eqnarray}

We use unitarity techniques to isolate the contribution of boxes~\cite{Britto:2004nc},
triangles~\cite{Forde:2007mi} and bubbles~\cite{Mastrolia:2009dr,Kilgore:2007qr,Davies:2011vt}.
Pentagon contributions to box coefficients are isolated by applying generalised
unitarity cuts on five propagators in $d=(4-2\epsilon)$-dimensions, with subsequent modification
as necessary to remove unphysical singularities and improve numerical stability~\cite{Budge:2020oyl}.
The coefficients of each integral are subsequently simplified using the techniques of momentum
twistors~\cite{Hodges:2009hk,Badger:2013gxa,Badger:2016uuq,Hartanto:2019uvl} and high precision
floating-point arithmetic~\cite{DeLaurentis:2019phz}.  We exploit previous results obtained in
the calculation of the same processes mediated by a fermion loop~\cite{Budge:2020oyl}, noting that
for our normalization the coefficients of bubble integrals, some triangle integrals, and the rational
part are identical.  The results of our analytic calculation of the amplitudes are presented in
full in section~\ref{sec:coeffs}.

\subsection{Squared matrix elements for fermion and scalar loops}
With the scalar-mediated amplitude calculations in hand, we can now describe the calculation of the
matrix elements relevant for the MSSM scenario described in Section~\ref{MSSMsection}.  For
simplicity and practicality we will include only a top-quark loop in the SM calculation, although
the inclusion of a bottom quark loop is straightforward.  We can write the subamplitude for a
$n-$parton process mediated by the top-quark, ${\tilde t}_1$ and ${\tilde t}_2$ as,
\begin{equation}
M^x_n = \left(\frac{m_t^2}{v}\right) H^x_n(m_t)
 - \left(\frac{\lambda_{h {\tilde t}_1 {\tilde t}_1}}{4}\right) A^x_n(m_{{\tilde t}_1})
 - \left(\frac{\lambda_{h {\tilde t}_2 {\tilde t}_2}}{4}\right) A^x_n(m_{{\tilde t}_2}) \, .
\end{equation}
Note that we have taken care to label the mass-dependence of the individual fermion and scalar subamplitudes,
and this formula applies to any of the subamplitudes, e.g. $x = 1234$ ($gggg$), $x=34$
($\bar q q gg$) or $x=4q$ ($\bar q q \bar q q$).  Expressions for all the relevant fermion-mediated
subamplitudes $H^x_n$ are given in Ref.~\cite{Budge:2020oyl}.

We can now form the squared matrix elements used in our calculation.  
For the four-gluon case we  can square the amplitude for a fixed helicity configuration and sum over
colours to find, 
\begin{eqnarray}
\label{coloursum4}
\sum_{\rm colours}\left|\cM^{gggg}_4\right|^2 &=& \bigg(\frac{g_s^4}{16 \pi^2}\bigg)^2 (N^2-1) 
\Bigg\{2 N^2 \big(\left|M^{1234}_4\right|^2+\left|M^{1342}_4\right|^2+\left|M^{1423}_4\right|^2\big) \nonumber \\
                   &-&4 \frac{(N^2-3)}{N^2} \left|M^{1234}_4+M^{1342}_4+M^{1423}_4\right|^2 \Bigg\}\, , 
\end{eqnarray}
where $N$ is the dimensionality of the $SU(N)$ colour group, i.e.~$N=3$, and the labels for the helicity configuration
(as explicitly shown in Eq.~(\ref{explicitfor4})) have been suppressed. 

Squaring the $\bar q q gg$ amplitude and summing over colours yields,
\begin{equation}
\sum |{\cM}^{{\qb}qgg}_4|^2 =\Big(\frac{g_s^4}{16 \pi^2 }\Big)^2
 \, (N^2-1) \, \left[ N \left( |M^{34}_{4}|^2 + |M^{43}_{4}|^2 \right)
  - \frac{1}{N} |M^{34}_{4}+M^{43}_{4}|^2 \right] \,,
\end{equation}
where the labelling of the helicity configuration shown in Eq.~(\ref{explicitfor2q}) has again been suppressed.

Squaring and summing the four-quark amplitude over colours gives,
\begin{equation}
\sum |{\cM}^{4q}_4(h_1, h_3)|^2 =\Big(\frac{g_s^4}{16 \pi^2 }\Big)^2
 \, (N^2-1) \, |M^{4q}_{4}(h_1,h_3)|^2 \, ,
\end{equation}
when the quark lines have different flavours.
For the case of identical quarks the sum over colours gives,
\begin{eqnarray}
\sum |{\cM}^{4q}_4|^2 &=& \Big(\frac{g_s^4}{16 \pi^2 }\Big)^2 \, (N^2-1)\, \Bigg( 
|M^{4q}_{4}(h_1,h_3)|^2 +|M^{4q^\prime}_{4}(h_1,h_3)|^2 \nonumber \\
 && \qquad + \frac{\delta_{h_1h_3}}{N} \left(M^{4q}_{4}(h_1,h_3) M^{4q^\prime}_{4}(h_1,h_3)^*
                                            +M^{4q}_{4}(h_1,h_3)^* M^{4q^\prime}_{4}(h_1,h_3) \right) \Bigg) \, ,
\end{eqnarray}
where, as indicated, the term on the second line only contributes for quarks of the same
helicity and we have introduced,
\begin{equation}
M^{4q^\prime}_{4}(h_1,h_3) = M^{4q}_{4}(1_{{\qb}}^{h_1},4_{q}^{-h_1},3_{{\qb}}^{h_3},2_{q}^{-h_3})\,.
\end{equation} 

\section{Four-parton integral coefficients for a scalar loop}
\label{sec:coeffs}

In this section we provide expressions for all the box, triangle and bubble
coefficients that can be treated as a minimal independent set for Higgs
plus four parton helicity amplitudes mediated by a massive scalar. 
Additional coefficients that can be obtained by momenta permutations 
and/or helicity flips are tabulated appropriately. 
As already noted, some of the coefficients 
are identical to the case when the circulating particle is a massive fermion~\cite{Budge:2020oyl}.
Specifically, these are:
\begin{itemize}
\item the entire rational contribution, $\tilde{r}$
\item bubble coefficients, $\tilde b_i$
\item a subset of triangle coefficients, $\tilde c_i$, where $i$ labels all triangles where none of the
external legs corresponds to the momentum of the Higgs boson
\item the $m^2$-dependent term in all triangle coefficients, $\tilde c_i^{(2)}$ (where, in general,
we expand $\tilde c_i = \tilde c_i^{(0)} + \tilde c_i^{(2)} m^2$)
\end{itemize} 
Expressions for such coefficients are not given here explicitly.  Instead we refer
to the equation numbers in ref.~\cite{Budge:2020oyl} where they have already been reported.
When the expressions for the whole coefficient coincide, these references are appended
(inside brackets) in subsequent tables.   For the partial coefficients,  $\tilde c_i^{(2)}$,
these references are included in the text.

As an aid to implementing these formulae in a numerical code, in appendix~\ref{sec:PScheck} we
provide values for all of the coefficients provided here, when evaluated at a specific phase
space point.  We also give the numerical values of the finite parts of the full amplitudes,
obtained by combining these coefficients with an evaluation of all loop integrals according
to Eq.~(\ref{scalarreduction4gluons}) (and its generalization to all partonic channels).

An additional check of the results presented here is that, in the limit of large mediator
mass, all the amplitudes should match onto limiting forms obtained by an explicit calculation
in the EFT.  This equivalence is spelled-out explicitly in appendix~\ref{sec:largemass}.

\subsection{Coefficients for $A_4^{1234}(g^{+},g^{+},g^{+},g^{+};\Higgs)$}
\label{pppp}
For the case with four gluons of positive helicity the complete result can easily be written in 
a form that includes a term proportional to the pentagon scalar integral ($E_0$),
\begin{eqnarray} \label{Higgs4gluons}
A_4 (\Higgs;1_g^+,2_g^+,3_g^+,4_g^+)&=&   \Bigg[\Bigg\{ \frac{4\,m^2}{\spa1.2\spa2.3\spa3.4\spa4.1} 
 \Big[ -\TrfourgR1.2.3.4  m^2 E_0(p_1,p_2,p_3,p_4;m)   \nonumber \\
&+& \frac{1}{2} ((s_{12}+s_{13})(s_{24}+s_{34})-s_{14}s_{23}) 
  D_0(p_1,p_{23},p_4;m)\nonumber \\
&+&\frac{1}{2} s_{12} s_{23} D_0(p_{1},p_{2},p_{3};m)\nonumber \\
&+& (s_{12}+s_{13}+s_{14}) C_0(p_1,p_{234};m)\Big] +2 \frac{s_{12}+s_{13}+s_{14}}{\spa1.2\spa2.3\spa3.4\spa4.1} 
\Bigg\}\nonumber \\
&+&\Bigg\{ 3~{\rm cyclic~permutations}\Bigg\}\Bigg] \, .
\end{eqnarray}
However, for consistency with the rest of our results and ease of implementation in a numerical code,
we prefer to present this result in terms of only box, triangle and bubble coefficients.  
The box coefficients then take a very simple form when written in terms of the 
effective pentagon coefficient (obtained by calculating the pentagon coefficient in $d$ dimensions
and taking the $\mu^2\to 0$ limit), 
\begin{eqnarray}
\etildehat_{1\x2\x3\x4} (1^+,2^+,3^+,4^+)=   -4\,m^4 \frac{\TrfourgR1.2.3.4 }{\spa1.2\spa2.3\spa3.4\spa4.1}
=-4\,m^4 \frac{\spb1.2\spb3.4 }{\spa1.2\spa3.4}\,.
\end{eqnarray}

The minimal set of integral coefficients needed to reconstruct the amplitude for this approach
is given in the first and third columns of Table~\ref{table:pppp}.

\begin{table}
\begin{center}
\begin{tabular}{|l|l||l|l|}
\hline
Coefficient          & Related coefficients &Coefficient          & Related coefficients\\
\hline
  $\tilde{d}_{1\x2\x34}$ & $\tilde{d}_{2\x3\x41},\tilde{d}_{3\x4\x12},\tilde{d}_{4\x1\x23},$   &   $\tilde{c}_{1\x234}$   & $\tilde{c}_{2\x341},\tilde{c}_{3\x412},\tilde{c}_{4\x123}$\\
                 & $\tilde{d}_{1\x4\x32},\tilde{d}_{2\x1\x43},\tilde{d}_{3\x2\x14},\tilde{d}_{4\x3\x21}$ &&\\
  $\tilde{d}_{1\x23\x4}$ & $\tilde{d}_{2\x34\x1},\tilde{d}_{3\x41\x2},\tilde{d}_{4\x12\x3}$ &&\\
  $\tilde{d}_{1\x2\x3}$  & $\tilde{d}_{2\x3\x4},\tilde{d}_{3\x4\x1},\tilde{d}_{4\x1\x2}$ && \\
\hline
\end{tabular}
\caption{Minimal set of integral coefficients for $A_4^{1234}(g^{+},g^{+},g^{+},g^{+};\Higgs)$.}
\label{table:pppp}
\end{center}
\end{table}

\subsubsection{$\tilde{d}_{1\x2\x34}$}
\begin{eqnarray}
\tilde{d}_{1\x2\x34}=\mC_{1\x2\x3\x4}^{(4)} \, \etildehat_{\{1^+\x2^+\x3^+\x4^+\}} \, .
\end{eqnarray}
\subsubsection{$\tilde{d}_{1\x23\x4}$}
\begin{eqnarray}
  \tilde{d}_{1\x23\x4}(1^+,2^+,3^+,4^+) &=&  \mC_{1\x2\x3\x4}^{(3)}\,\etildehat_{\{1^+\x2^+\x3^+\x4^+\}} \nonumber \\
                                &+&\frac{2\,m^2}{\spa1.2\spa2.3\spa3.4\spa4.1}\,[(s_{12}+s_{13})\,(s_{24}+s_{34})-s_{14}\,s_{23}] \, .
\end{eqnarray}
\subsubsection{$\tilde{d}_{1\x2\x3}$}
\begin{eqnarray}
    \tilde{d}_{1\x2\x3}&=& \mC_{4\x1\x2\x3}^{(1)}\,\etildehat_{\{4^+\x1^+\x2^+\x3^+\}} + \mC_{1\x2\x3\x4}^{(5)}\,\etildehat_{\{1^+\x2^+\x3^+\x4^+\}}\nonumber  \\
                &+& \frac{2\,m^2}{\spa1.2\spa2.3\spa3.4\spa4.1} \,s_{12}\,s_{23} \, .
\end{eqnarray}
\subsubsection{$\tilde{c}_{1\x234}^{(0)}$, $\tilde{c}_{1\x234}^{(2)}$}
\begin{eqnarray}
\tilde{c}_{1\x234}^{(0)}(1^+,2^+,3^+,4^+)=0 \, .
\end{eqnarray}
We have $\tilde{c}_{1\x234}^{(2)}(1^+,2^+,3^+,4^+)=c_{1\x234}^{(2)}(1^+,2^+,3^+,4^+)$ where the fermionic coefficient is given in 
Eq.~(4.10) of ref.~\cite{Budge:2020oyl}.


\subsection{Coefficients for $A_4^{1234}(g^{+},g^{+},g^{+},g^{-};\Higgs)$}
\label{pppm}

The effective pentagon coefficients~\cite{Budge:2020oyl} used to define the box coefficients below
are,
\begin{eqnarray}
\etildehat_{\{1^+\x2^+\x3^+\x4^-\}} &=& -4\,m^4 \frac{\spb2.3 \, \spab4.(2+3).1}{\spa2.3 \, \spab1.(2+3).4} \, , \\
\etildehat_{\{4^-\x1^+\x2^+\x3^+\}} &=& \etildehat_{\{1^+\x2^+\x3^+\x4^-\}}\{1\leftrightarrow3\}\, , \\
\etildehat_{\{2^+\x3^+\x4^-\x1^+\}} &=& -4\,m^4
 \frac{\spb2.3^2 \spa3.4 \spab2.(3+4).1}{\spa2.3^2 \spb3.4 \spab1.(3+4).2} \, ,\\
\etildehat_{\{3^+\x4^-\x1^+\x2^+\}} &=& \etildehat_{\{2^+\x3^+\x4^-\x1^+\}}\{1\leftrightarrow3\}\, .
\end{eqnarray}

The minimal set of integral coefficients needed to reconstruct the amplitude for this approach
is given in the first and third columns of Table~\ref{table:pppm}.

\begin{table}
\begin{center}
\begin{tabular}{|l|l||l|l|}
\hline
Coefficient          & Related coefficients & Coefficient          & Related coefficients \\
\hline
  $\tilde{d}_{1\x2\x34}$ & $\tilde{d}_{3\x2\x14}$&                             $ \tilde{c}_{3\x4}$  (5.19) & $\tilde{c}_{4\x1}$  \\  
  $\tilde{d}_{1\x4\x32}$ & $\tilde{d}_{3\x4\x12}$&                             $ \tilde{c}_{2\x34}$ (5.20) & $\tilde{c}_{2\x14}$ \\  
  $\tilde{d}_{2\x1\x43}$ & $\tilde{d}_{2\x3\x41}$&                             $ \tilde{c}_{1\x43}$ (5.21) & $\tilde{c}_{3\x41}$ \\  
  $\tilde{d}_{4\x3\x21}$ & $\tilde{d}_{4\x1\x23}$&                             $ \tilde{c}_{4\x123}$       & \\              
  $\tilde{d}_{2\x34\x1}$ & $\tilde{d}_{3\x41\x2}$&                             $ \tilde{c}_{1\x234}$       & $\tilde{c}_{3\x412}$ \\ 
  $\tilde{d}_{1\x23\x4}$ & $\tilde{d}_{4\x12\x3}$&                             $ \tilde{c}_{2\x341}$       & \\               
  $\tilde{d}_{2\x3\x4}$  & $\tilde{d}_{4\x1\x2}$ &                             $ \tilde{c}_{12\x34}$       & $\tilde{c}_{23\x41}$ \\  
  $\tilde{d}_{1\x2\x3}$  &                       &                             $ \tilde{b}_{34}$     (5.30)& $\tilde{b}_{14}$ \\            
  $\tilde{d}_{3\x4\x1}$  &                       &                             $ \tilde{b}_{234}$    (5.31)& $\tilde{b}_{412},\tilde{b}_{341}$\\ 
                         &                       &                             $ \tilde{b}_{1234}$   (5.32)& \\ 
\hline
\end{tabular}
\caption{Minimal set of integral coefficients for $A_4^{1234}(g^{+},g^{+},g^{+},g^{-};\Higgs)$.
The equation numbers in brackets 
give the place in ref.~\cite{Budge:2020oyl} where the coefficients are reported. These coefficients are 
the same in the scalar-mediated and the fermion-mediated theories.}
\label{table:pppm}
\end{center}
\end{table}

\subsubsection{$\tilde{d}_{1\x2\x34}$}
\begin{eqnarray}
  \tilde{d}_{1\x2\x34}(1^+,2^+,3^+,4^-) &=& \mC_{1\x2\x3\x4}^{(4)}\,\etildehat_{\{1^+\x2^+\x3^+\x4^-\}} 
  - 2\,m^2 \frac{\spb1.2}{\spa1.2} \Bigg[ \frac{\spa2.4^2\,\spab4.(2+3).1}{\spa2.3\,\spa3.4\,\spab2.(3+4).1} \nonumber \\
               &+&\frac{\spb2.3\,\spab1.(2+4).3^2}{\spb3.4\,\spab1.(3+4).2\,\spab1.(2+3).4} \Bigg] \, .
\end{eqnarray} \subsubsection{$\tilde{d}_{1\x4\x32}$}
\begin{eqnarray}
 \tilde{d}_{1\x4\x32}(1^+,2^+,3^+,4^-) = \mC_{2\x3\x4\x1}^{(2)}\,\etildehat_{\{2^+\x3^+\x4^-\x1^+\}} 
                              + 2\,m^2 \frac{\spb2.3s_{14}s_{234}^2}{\spa2.3^2\spb3.4\spab1.(3+4).2\spab1.(2+3).4} \, . \nonumber \\
\end{eqnarray} \subsubsection{$\tilde{d}_{2\x1\x43}$}
\begin{eqnarray}
  \tilde{d}_{2\x1\x43}(1^+,2^+,3^+,4^-) &=& \mC_{3\x4\x1\x2}^{(2)}\,\etildehat_{\{3^+\x4^-\x1^+\x2^+\}} 
   + 2\,m^2 \frac{\spb1.2}{\spa1.2} \Bigg[ \frac{\spb1.3^2\,\spab2.(1+4).3}{\spb1.4\,\spb3.4\,\spab2.(3+4).1} \nonumber \\
  &+& \frac{\spa1.4\,\spab4.(1+3).2^2}{\spa3.4\,\spab1.(3+4).2\,\spab3.(1+4).2} \Bigg] \, .
\end{eqnarray} \subsubsection{$\tilde{d}_{4\x3\x21}$}
\begin{eqnarray}
  \tilde{d}_{4\x3\x21}(1^+,2^+,3^+,4^-) = \mC_{1\x2\x3\x4}^{(2)} \, \etildehat_{\{1^+\x2^+\x3^+\x4^-\}} 
                                + 2\,m^2 \frac{s_{34}\,s_{123}^2}{\spa1.2\,\spa2.3\,\spab1.(2+3).4\,\spab3.(1+2).4} \, . \nonumber \\
\end{eqnarray} 
\subsubsection{$\tilde{d}_{1\x23\x4}$}
\begin{eqnarray}
    \tilde{d}_{1\x23\x4}(1^+,2^+,3^+,4^-) =  \mC_{1\x2\x3\x4}^{(3)} \, \etildehat_{\{1^+\x2^+\x3^+\x4^-\}} \, .
\end{eqnarray} \subsubsection{$\tilde{d}_{2\x34\x1}$}
\begin{eqnarray}
  \tilde{d}_{2\x34\x1}(1^+,2^+,3^+,4^-) &=& \mC_{2\x3\x4\x1}^{(3)}\,\etildehat_{\{2^+\x3^+\x4^-\x1^+\}} \nonumber \\ 
 &+& \frac{2\,\spa2.4}{\spa1.2\,\spa2.3} \Biggl[
 \frac{\spa1.4\,\spa2.4\,\spab1.(3+4).2\,\spab2.(3+4).1}{\spa1.2^2\,\spa3.4} \nonumber \\ 
 &-& m^2 \left( 3\,\frac{\spa1.4\,\spa2.4\,\spb1.2}{\spa1.2\,\spa3.4} 
                  +2 \frac{\spb1.3\,\spb2.3}{\spb3.4}
                  +\frac{\spa2.4\,\spb1.4\,\spb2.3}{\spa2.3\,\spb3.4} \right) \Biggr] \, .
\end{eqnarray} 
\subsubsection{$\tilde{d}_{2\x3\x4}$}
\begin{eqnarray}
  \tilde{d}_{2\x3\x4}(1^+,2^+,3^+,4^-)  &=& \mC_{1\x2\x3\x4}^{(1)}\,\etildehat_{\{1^+\x2^+\x3^+\x4^-\}} + \mC_{2\x3\x4\x1}^{(5)}\,\etildehat_{\{2^+\x3^+\x4^-\x1^+\}}\nonumber\\
                                &+& 2\,m^2 \frac{s_{234}\,\spa3.4\,\spb2.3^2}{\spa2.3\,\spab1.(3+4).2\,\spab1.(2+3).4}	\, .			
\end{eqnarray} \subsubsection{$\tilde{d}_{1\x2\x3}$}
\begin{eqnarray}
  \tilde{d}_{1\x2\x3}(1^+,2^+,3^+,4^-) &=& \mC_{1\x2\x3\x4}^{(5)} \, \etildehat_{\{1^+\x2^+\x3^+\x4^-\}} + \mC_{4\x1\x2\x3}^{(1)} \, \etildehat_{\{4^-\x1^+\x2^+\x3^+\}} \nonumber \\
                               &+& 2\,m^2 \frac{s_{123}\,\spb1.2\,\spb2.3}{\spab3.(1+2).4\,\spab1.(2+3).4} \, .
\end{eqnarray} \subsubsection{$\tilde{d}_{3\x4\x1}$}
\begin{eqnarray}
  \tilde{d}_{3\x4\x1}(1^+,2^+,3^+,4^-) &=& \mC_{2\x3\x4\x1}^{(1)}\,\etildehat_{\{2^+\x3^+\x4^-\x1^+\}} + \mC_{3\x4\x1\x2}^{(5)}\,\etildehat_{\{3^+\x4^-\x1^+\x2^+\}} \nonumber \\
                               &-& {2 m^2} \frac{1}{\spa1.3} \Bigg[ \frac{\spb2.3\,\spa3.4^2\,s_{14}}{\spa2.3^2\,\spab1.(3+4).2} 
                                +   \frac{\spb1.2\,\spa1.4^2\,s_{34}}{\spa1.2^2\,\spab3.(1+4).2} \Bigg] \nonumber  \\
                               &+& \frac{\spa1.4\,\spa3.4}{\spa1.2\,\spa1.3^2\,\spa2.3}\Bigg[2\,s_{14}\,s_{34} + 6\,m^2\,s_{13}\Bigg]\,. 
\end{eqnarray}

\subsubsection{$\tilde{c}_{4\x123}^{(0)}$, $\tilde{c}_{4\x123}^{(2)}$}
\begin{equation}
\tilde{c}_{4\x123}^{(0)}(1^+,2^+,3^+,4^-)=0 \, .
\end{equation}
We have 
$\tilde{c}_{4\x123}^{(2)}(1^+,2^+,3^+,4^-) = c_{4\x123}^{(2)}(1^+,2^+,3^+,4^-)$,
where the fermionic coefficient is given in Eq.~(5.23) of ref.~\cite{Budge:2020oyl}.

\subsubsection{$\tilde{c}^{(0)}_{1\x234}$, $\tilde{c}^{(2)}_{1\x234}$}
\begin{eqnarray}
\tilde{c}_{1\x234}^{(0)}(1^+,2^+,3^+,4^-)&=& 2 (s_{12}+s_{13}+s_{14}) 
\frac{\spa1.4 \,\spa2.4^2}{\spa1.2^3 \,\spa2.3 \,\spa3.4 } \, . \nonumber \\
\end{eqnarray}
Also, we have 
$\tilde{c}_{1\x234}^{(2)}(1^+,2^+,3^+,4^-) = c_{1\x234}^{(2)}(1^+,2^+,3^+,4^-)$,
where the fermionic coefficient is given in Eq.~(5.25) of ref.~\cite{Budge:2020oyl}.

\subsubsection{$\tilde{c}^{(0)}_{2\x341}$, $\tilde{c}_{2\x341}^{(2)}$}
\begin{eqnarray}
\tilde{c}_{2\x341}^{(0)}(1^+,2^+,3^+,4^-)&=& 2 (s_{12}+s_{23}+s_{24}) \, \spa2.4^2 \,
\frac{\spa1.4^2 \,\spa2.3^2 + \spa1.2^2\,\spa3.4^2}{\spa1.2^3 \,\spa2.3^3 \,\spa1.4 \,\spa3.4 } \, .
\nonumber \\
\end{eqnarray}
Furthermore, 
$\tilde{c}_{2\x341}^{(2)}(1^+,2^+,3^+,4^-) = c_{2\x341}^{(2)}(1^+,2^+,3^+,4^-)$,
where the fermionic coefficient is given in Eq.~(5.27) of ref.~\cite{Budge:2020oyl}.

\subsubsection{$\tilde{c}^{(0)}_{12\x34}$, $\tilde{c}^{(2)}_{12\x34}$}
\begin{eqnarray}
\tilde{c}_{12\x34}^{(0)}(1^+,2^+,3^+,4^-)
 &=&  0 \, .
\end{eqnarray}
In addition, 
$\tilde{c}_{12\x34}^{(2)}(1^+,2^+,3^+,4^-) = c_{12\x34}^{(2)}(1^+,2^+,3^+,4^-)$,
where the fermionic coefficient is given in Eq.~(5.29) of ref.~\cite{Budge:2020oyl}.

\subsection{Coefficients for $A_4^{1234}(g^{+},g^{-},g^{+},g^{-};\Higgs)$}
\label{pmpm}

The effective pentagon coefficients for this helicity combination are,
\begin{eqnarray}
\etildehat_{\{1^+\x2^-\x3^+\x4^-\}}&=& -4 m^4 \frac{\spa1.2\spb3.4\spab4.(2+3).1^2}{\spb1.2\spa3.4\spab1.(2+3).4^2} \, ,\\
\etildehat_{\{3^+\x4^-\x1^+\x2^-\}}&=& \etildehat_{\{1^+\x2^-\x3^+\x4^-\}}
 \Bigl\{ 1 \leftrightarrow 3, \, 2 \leftrightarrow 4 \Bigr\} \, , \\
\etildehat_{\{4^-\x1^+\x2^-\x3^+\}}&=& \etildehat_{\{1^+\x2^-\x3^+\x4^-\}}
 \Bigl\{ 1 \to 4, \, 2 \to 1, \, 3 \to 2, \, 4 \to 3, \,
 \langle \rangle \leftrightarrow [] \Bigr\} \, , \\
\etildehat_{\{2^-\x3^+\x4^-\x1^+\}}&=& \etildehat_{\{1^+\x2^-\x3^+\x4^-\}}
 \Bigl\{ 1 \to 2, \, 2 \to 3, \, 3 \to 4, \, 4 \to 1, \,
 \langle \rangle \leftrightarrow [] \Bigr\} \, .
\end{eqnarray}

The minimal set of coefficients that needs to be calculated is given in Table~\ref{table:pmpm}.

\begin{table}
\begin{center}
\begin{tabular}{|l|l||l|l|}
\hline
Coefficient      & Related coefficients                                    &   Coefficient    & Related coefficients \\
\hline
  $\tilde{d}_{4\x3\x21}$ & $\tilde{d}_{2\x1\x43},\tilde{d}_{3\x2\x14},\tilde{d}_{1\x4\x32},$               &   $\tilde{c}_{3\x4} $ (6.9)   & $\tilde{c}_{4\x1},\tilde{c}_{2\x3},\tilde{c}_{1\x2}$\\               
                 & $\tilde{d}_{1\x2\x34},\tilde{d}_{2\x3\x41},$                                            &   $\tilde{c}_{2\x34}$ (6.10 ) & $\tilde{c}_{3\x41},\tilde{c}_{4\x12},\tilde{c}_{1\x23}$\\            
                 & $\tilde{d}_{3\x4\x12},\tilde{d}_{4\x1\x23}$                                             &                               & $\tilde{c}_{1\x43},\tilde{c}_{2\x14},\tilde{c}_{3\x21},\tilde{c}_{4\x32}$\\ 
 $\tilde{d}_{1\x23\x4}$ & $\tilde{d}_{2\x34\x1},\tilde{d}_{3\x41\x2},\tilde{d}_{4\x12\x3}$                 &   $\tilde{c}_{12\x34}$   & $\tilde{c}_{23\x41}$ \\                              
  $\tilde{d}_{1\x2\x3}$  & $\tilde{d}_{2\x3\x4},\tilde{d}_{3\x4\x1},\tilde{d}_{4\x1\x2}$                   &   $\tilde{c}_{1\x234}$   & $\tilde{c}_{2\x341},\tilde{c}_{3\x412},\tilde{c}_{4\x123}$\\         
                 &                                                         &   $\tilde{b}_{34}$  (6.17)    & $\tilde{b}_{12},\tilde{b}_{23},\tilde{b}_{41}$ \\                    
                 &                                                         &   $\tilde{b}_{234}$ (6.18)    & $\tilde{b}_{341},\tilde{b}_{412},\tilde{b}_{123}$ \\                 
                 &                                                         &   $\tilde{b}_{1234}$(6.19)    &  \\                 
\hline
\end{tabular}
\caption{Minimal set of integral coefficients for $A_4^{1234}(g^{+},g^{-},g^{+},g^{-};\Higgs)$.
The equation numbers in brackets
give the place in ref.~\cite{Budge:2020oyl} where the coefficients are reported. These coefficients are
the same in the scalar-mediated and the fermion-mediated theories.}
\label{table:pmpm}
\end{center}
\end{table}

\subsubsection{$\tilde{d}_{4\x3\x21}$}
\begin{eqnarray}
  \tilde{d}_{4\x3\x21}(1^+,2^-,3^+,4^-) &=& \etildehat_{\{1^+\x2^-\x3^+\x4^-\}} \,\mC_{1\x2\x3\x4}^{(2)} \nonumber \\
  &-&\frac{2\,\spab2.(1+3).4}{\spab1.(2+3).4\spab3.(1+2).4} \Bigg[
  \frac{\spa2.3\spab2.(1+3).4s_{34}s_{123}^2}{\spa1.2\spab3.(1+2).4^2} \nonumber \\
  &+& m^2 \Bigg(2\frac{\spb1.3\spab4.(2+3).1}{\spb1.2} + \frac{\spb2.3\spab2.(1+3).4\spab4.(2+3).1}{\spb1.2\spab1.(2+3).4} \nonumber \\
  &+& 3\frac{\spa2.3\spab2.(1+3).4\spab4.(1+2).3}{\spa1.2\spab3.(1+2).4} \Bigg) \Bigg] \, .
\end{eqnarray} \subsubsection{$\tilde{d}_{1\x23\x4}$}
\begin{eqnarray}
  \tilde{d}_{1\x23\x4}(1^+,2^-,3^+,4^-) &=& \etildehat_{\{1^+\x2^-\x3^+\x4^-\}} \,\mC_{1\x2\x3\x4}^{(3)} \nonumber \\
                                &-& 2\,m^2 \frac{\spab4.(2+3).1}{\spab1.(2+3).4} 
                                 \Bigg[ \frac{\spa1.2\spa2.4^2}{\spa1.4\spa2.3\spa3.4} + \frac{\spb1.3^2\spb3.4}{\spb1.2\spb1.4\spb2.3}\Bigg]  \, .
\end{eqnarray} \subsubsection{$\tilde{d}_{1\x2\x3}$}
\begin{eqnarray}
  \tilde{d}_{1\x2\x3}(1^+,2^-,3^+,4^-) &=& \mC_{1\x2\x3\x4}^{(5)} \,
                                   \etildehat_{\{1^+\x2^-\x3^+\x4^-\}} +
                                   \mC_{4\x1\x2\x3}^{(1)} \, \etildehat_ {\{4^-\x1^+\x2^-\x3^+\}} \nonumber \\
  &+&\frac{\spa1.2\spa2.3}{\spab1.(2+3).4\spab3.(1+2).4} \Bigg[ 
      -2\frac{s_{12}s_{23}s_{123}}{\spa1.3^2} + 2m^2 \Bigg(2\frac{\spb1.3s_{123}}{\spa1.3} \nonumber \\
  &-&\spb1.3^2 + \frac{\spb1.2\spb2.3\spa2.4^2}{\spa1.4\spa3.4} - \frac{\spb2.3\spab2.(1+3).4\spab4.(2+3).1}{\spa3.4\spab1.(2+3).4}  \nonumber \\
  &+& \frac{\spb1.2\spab2.(1+3).4\spab4.(1+2).3}{\spa1.4\spab3.(1+2).4} \Bigg) \Bigg] \, .
\end{eqnarray} 
\subsubsection{$\tilde{c}^{(0)}_{12\x34}$, $\tilde{c}^{(2)}_{12\x34}$}
In this case the scalar coefficient
$\tilde{c}_{12\x34}^{(0)}(1^+,2^-,3^+,4^-)$ has been given previously,
in Eq.~(6.13) of Ref.~\cite{Budge:2020oyl}.
Moreover, 
$\tilde{c}_{12\x34}^{(2)}(1^+,2^-,3^+,4^-) = c_{12\x34}^{(2)}(1^+,2^-,3^+,4^-)$,
where the fermionic coefficient is given in Eq.~(6.12) of ref.~\cite{Budge:2020oyl}.

 \subsubsection{$\tilde{c}_{1\x234}^{(0)}$, $\tilde{c}_{1\x234}^{(2)}$}

\begin{eqnarray}
\tilde{c}_{1\x234}^{(0)}(1^+,2^-,3^+,4^-)&=&
 -2 (s_{12}+s_{13}+s_{14})\frac{s_{234}\, \spab1.(2+4).3^2 }
 {\spb2.3 \, \spb3.4 \, \spab1.(2+3).4^3   \, \spab1.(3+4).2^3}
\nonumber \\
&\times & \Big(\spb2.3^2\,\spab1.(2+3).4^2
           +    \spb3.4^2\,\spab1.(3+4).2^2 \Big) \, . \nonumber \\ 
\end{eqnarray}
In addition, we have 
$\tilde{c}_{1\x234}^{(2)}(1^+,2^-,3^+,4^-) = c_{1\x234}^{(2)}(1^+,2^-,3^+,4^-)$,
where the fermionic coefficient is given in Eq.~(6.16) of ref.~\cite{Budge:2020oyl}.

\subsection{Coefficients for $A_4^{1234}(g^{+},g^{+},g^{-},g^{-};\Higgs)$}
\label{ppmm}

The effective pentagon coefficients are given by,
\begin{eqnarray}
\etildehat_{\{1^+\x2^+\x3^-\x4^-\}}
  &=& -4\,m^4
\frac{\spb1.2 \spa3.4}{\spa1.2 \spb3.4 } \, , \\
\etildehat_{\{2^+\x3^-\x4^-\x1^+\}} &=& -4\, m^4 \frac{\spb2.3\,\spa3.4^2\,\spb4.1}{\spa2.3\,\spb3.4^2\,\spa4.1} \, ,  \\
\etildehat_{\{2^+\x3^-\x4^-\x1^+\}} &=& -4\,m^4 \frac{\spb2.3\,\spa3.4^2\,\spb4.1}{\spa2.3\,\spb3.4^2\,\spa4.1} \, , \\
\etildehat_{\{4^-\x1^+\x2^+\x3^-\}} &=& \etildehat_{\{2^+\x3^-\x4^-\x1^+\}}\{2\leftrightarrow4, 1\leftrightarrow3, \langle\,\rangle\leftrightarrow[\,]\} \, .
\end{eqnarray}

The minimal set of coefficients that needs to be calculated is given in Table~\ref{table:ppmm}.

\begin{table}
\begin{center}
\begin{tabular}{|l|l||l|l|}
\hline
Coefficient          & Related coefficients                     &Coefficient          & Related coefficients \\
\hline
  $\tilde{d}_{1\x2\x34}$ & $\tilde{d}_{2\x1\x43},\tilde{d}_{3\x4\x12},\tilde{d}_{4\x3\x21} $    &  $\tilde{c}_{2\x3}$ (7.10)    & $\tilde{c}_{4\x1}$\\                           
  $\tilde{d}_{1\x4\x32}$ & $\tilde{d}_{3\x2\x14},\tilde{d}_{4\x1\x23},\tilde{d}_{2\x3\x41} $    &  $\tilde{c}_{1\x23}$ (7.11)   & $\tilde{c}_{2\x14},\tilde{c}_{3\x41},\tilde{c}_{4\x32}$  \\    
  $\tilde{d}_{2\x34\x1}$ & $\tilde{d}_{4\x12\x3}$                               &  $\tilde{c}_{1\x234}$   & $\tilde{c}_{2\x341},\tilde{c}_{3\x412} ,\tilde{c}_{4\x123}$ \\                        
  $\tilde{d}_{1\x23\x4}$ & $\tilde{d}_{3\x41\x2}$                               &  $\tilde{c}_{23\x41}$   &  \\ 
  $\tilde{d}_{1\x2\x3}$  & $\tilde{d}_{3\x4\x1},\tilde{d}_{4\x1\x2},\tilde{d}_{2\x3\x4}$        &  $\tilde{b}_{23}$  (7.17)     & $\tilde{b}_{41}$ \\                            
                 &                                              &  $\tilde{b}_{234}$  (7.18)    & $\tilde{b}_{341},\tilde{b}_{412},\tilde{b}_{123}$ \\           
                 &                                              &  $\tilde{b}_{1234}$ (7.19)    & \\           
\hline
\end{tabular}
\caption{Minimal set of integral coefficients for $A_4^{1234}(g^{+},g^{+},g^{-},g^{-};\Higgs)$. 
The equation numbers in brackets 
give the place in ref.~\cite{Budge:2020oyl} where the coefficients are reported. These coefficients are 
the same in the scalar-mediated and the fermion-mediated theories.}
\label{table:ppmm}
\end{center}
\end{table}

\subsubsection{$\tilde{d}_{1\x2\x34}$}
\begin{eqnarray}
    \tilde{d}_{1\x2\x34}(1^+,2^+,3^-,4^-)= \mC_{1\x2\x3\x4}^{(4)} \, \etildehat_{\{1^+\x2^+\x3^-\x4^-\}} \, .
\end{eqnarray}

 \subsubsection{$\tilde{d}_{1\x4\x32}$}
\begin{eqnarray}
  \tilde{d}_{1\x4\x32}(1^+,2^+,3^-,4^-)&=& \mC_{2\x3\x4\x1}^{(2)} \,\etildehat_ {\{2^+\x3^-\x4^-\x1^+\}} 
  - 2\, \frac{\spb2.4^2}{\spab1.(2+3).4\,\spb3.4}
  \Bigg\{ \frac{s_{14}\,s_{234}^2 \spab1.(3+4).2}{\spb2.3\,\spab1.(2+3).4^2}  \nonumber \\
  &+& m^2 \Big[3\,\frac{\spab1.(3+4).2\spab4.(2+3).1}{\spb2.3\spab1.(2+3).4}
  +\frac{\spb1.4\spa3.4s_{234}}{\spa2.3\spb2.4\spb3.4} 
  +\frac{\spa3.4\,\spab3.(2+4).1}{\spa2.3\spb2.4}\Big]\Bigg\} \, . \nonumber \\
\end{eqnarray}      
 
\subsubsection{$\tilde{d}_{2\x34\x1}$}
\begin{eqnarray}
  \tilde{d}_{2\x34\x1}(1^+,2^+,3^-,4^-)&=& \mC_{2\x3\x4\x1}^{(3)} \,\etildehat_{\{2^+\x3^-\x4^-\x1^+\}}
  -2\, m^2 \frac{\spa3.4\,\spab1.(3+4).2\,\spab2.(3+4).1}{\spa1.2\,\spa1.4\,\spa2.3\,\spb3.4^2} \, .
  \nonumber \\
\end{eqnarray}
 \subsubsection{$\tilde{d}_{1\x23\x4}$}
\begin{eqnarray}
\tilde{d}_{1\x23\x4}(1^+,2^+,3^-,4^-)&=& \mC_{1\x2\x3\x4}^{(3)}   \, \etildehat_{\{1^+\x2^+\x3^-\x4^-\}}  \nonumber \\
&+& 2\,m^2\,\frac{\spab4.(2+3).1}{\spa1.2\,\spb3.4\,\spab1.(2+3).4} \times 
\Bigg[ \frac{s_{12} \spb2.4^2}{\spb1.4\,\spb2.3} + \frac{s_{34} \spa1.3^2}{\spa2.3\,\spa1.4} \Bigg] \, .
\end{eqnarray}
 
\subsubsection{$\tilde{d}_{1\x2\x3}$}
\begin{eqnarray}
\tilde{d}_{1\x2\x3}(1^+,2^+,3^-,4^-)&=& \mC_{1\x2\x3\x4}^{(5)} \,\etildehat_{\{1^+\x2^+\x3^-\x4^-\}}
               +\mC_{4\x1\x2\x3}^{(1)}\,\etildehat_{\{4^-\x1^+\x2^+\x3^-\}} \nonumber \\
&-& 2\, m^2 \frac{\spb1.2^2\,\spa2.3}{\spa1.2\,\spb1.4\,\spb3.4} \, .
\end{eqnarray}
 
\subsubsection{$\tilde{c}^{(0)}_{23\x41}$, $\tilde{c}^{(2)}_{23\x41}$}
\begin{eqnarray}
\tilde{c}_{23\x41}^{(0)}(1^+,2^+,3^-,4^-)= -\tilde{c}_{12\x34}^{(0)}(2^+,3^-,1^+,4^-) 
&-& \symbrack\Bigg\{ 2\,\DeltaThree(1,4,2,3) \Big[\frac{(s_{13}-s_{24})}{\spab2.(1+4).3\,\spab1.(2+3).4}\Big]^2 \nonumber \\
    &+&4\frac{\spab3.(1+4).2\,\spab4.(2+3).1}{\spab2.(1+4).3\,\spab1.(2+3).4}\symbrack\Bigg\} \, . \nonumber \\
\end{eqnarray}
Moreover, 
$\tilde{c}_{23\x41}^{(2)}(1^+,2^+,3^-,4^-) = c_{23\x41}^{(2)}(1^+,2^+,3^-,4^-)$,
where the fermionic coefficient is given in Eq.~(7.14) of ref.~\cite{Budge:2020oyl}.

 \subsubsection{$\tilde{c}^{(0)}_{1\x234}$, $\tilde{c}^{(2)}_{1\x234}$}
\begin{eqnarray}
  \tilde{c}_{1\x234}^{(0)}(1^+,2^+,3^-,4^-) &=&
  -2 \, (s_{12}+s_{13}+s_{14}) \, s_{234} \,
 \frac{\spab1.(3+4).2 \spb2.4^2}{\spab1.(2+3).4^3 \spb2.3 \spb3.4} \, .
\end{eqnarray}
Furthermore, 
$\tilde{c}_{1\x234}^{(2)}(1^+,2^+,3^-,4^-) = c_{1\x234}^{(2)}(1^+,2^+,3^-,4^-)$,
where the fermionic coefficient is given in Eq.~(7.16) of ref.~\cite{Budge:2020oyl}.

\subsection{Coefficients for $A^{34}_4(\qb^+,q^-,g^+,g^+;\Higgs)$}

The coefficients that must be computed for this amplitude are shown in
the left-hand column of Table~\ref{table:aqgg}.

\renewcommand{\baselinestretch}{1.2}
\begin{table}
\begin{center}
\begin{tabular}{|l|l||l|l||l|}
\hline
\multicolumn{2}{|c||}{$1_{\qb}^+,2_q^-,3_g^+,4_g^+$}     & \multicolumn{2}{|c||}{$1_{\qb}^+,2_q^-,3_g^-,4_g^+$}  & \multicolumn{1}{|c|}{$1_{\qb}^+,2_q^-,3_g^+,4_g^-$}   \\
\hline
Coefficient          & Related coefficient   & Coefficient          & Related coefficient  & Coefficient  \\
\hline
$\tilde d_{3\x21\x4}$ &                  &$\tilde d_{3\x21\x4}$ & & $\tilde c_{4\x123}$\\                            
$\tilde d_{4\x3\x21}$ & $\tilde d_{3\x4\x12}$   &$\tilde d_{4\x3\x21}$ & $\tilde d_{3\x4\x12}$ & $\tilde b_{123}$(10.4)\\             
$\tilde c_{3\x21}$ (8.4) & $\tilde c_{4\x12}$   &$\tilde c_{3\x4}$ (9.3)& & \\                         
$\tilde c_{12\x34}$ &                    &$\tilde c_{3\x21}$ (9.4) & $\tilde c_{4\x12}$ & \\                                
$\tilde c_{4\x123}$ &                    &$\tilde c_{12\x34}$ & & \\                              
$\tilde c_{3\x412}$ &                    &$\tilde c_{4\x123}$ & $\tilde c_{3\x412}$ & \\                 
$\tilde b_{12}$ (8.11) &                        &$\tilde b_{34}$ (9.9)& & \\                                  
$\tilde b_{123}$ (8.12) &                       &$\tilde b_{12}$ (9.10)& & \\                                  
$\tilde b_{412}$ (8.13)&                       &$\tilde b_{123}$ (9.11)& $\tilde b_{412}$ & \\                       
$\tilde b_{1234}$ (8.14)&                      &$\tilde b_{1234}$ (9.12) & & \\                                                   
\hline
\end{tabular}
\caption{Minimal set of integral coefficients for $A^{34}_4(1_{\qb}^+,2_q^-,3_g^+,4_g^+)$, 
$A^{34}_4(1_{\qb}^+,2_q^-,3_g^-,4_g^+)$ and $A^{34}_4(1_{\qb}^+,2_q^-,3_g^+,4_g^-)$ together 
with the related coefficients that can be obtained from the base set.
The equation numbers in brackets give the place in ref.~\cite{Budge:2020oyl} where the coefficients are reported. These coefficients are 
the same in the scalar-mediated and the fermion-mediated theories.}
\label{table:aqgg}
\end{center}
\end{table}
\renewcommand{\baselinestretch}{1}

\subsubsection{$\tilde d_{3\x21\x4}$}
\begin{eqnarray}
\tilde d_{3\x21\x4}(1^+_{\bar{q}},2^-_q,3^+_g,4^+_g)&=&
      -2\,\frac{\spa2.4\,\spa2.3 }
   {\spa1.2\,\spa3.4^3} \, \Big[(s_{13}+s_{23})\,(s_{14}+s_{24})-s_{12}\,s_{34}\Big] \nonumber \\
   &+&2\, m^2 \Bigg[\frac{\spb1.3\,\spb1.4}{\spb1.2\,\spa3.4} 
    +3\,\frac{\spa2.3\,\spa2.4\,\spb3.4}{\spa1.2\,\spa3.4^2} \Bigg] \, .
\end{eqnarray}

\subsubsection{$\tilde d_{4\x3\x21}$}
\begin{eqnarray}
\tilde d_{4\x3\x21}(1^+_{\bar{q}},2^-_q,3^+_g,4^+_g)&=&
  2\, m^2\,\frac{\spb3.4}{\spa3.4}\,\Bigg[\frac{\spa2.3\,\spab2.(1+3).4}{\spa1.2\,\spab3.(1+2).4}
    -\frac{\spb1.3\,\spab4.(2+3).1}{\spb1.2\,\spab4.(1+2).3}\Bigg] \, .
\end{eqnarray}

\subsubsection{$\tilde c_{12\x34}^{(0)},\tilde c_{12\x34}^{(2)}$}
\begin{eqnarray}
\tilde c_{12\x34}^{(0)}(1^+_{\bar{q}},2^-_q,3^+_g,4^+_g)&=& 0 \, .
\end{eqnarray}
The coefficient $\tilde c_{12\x34}^{(2)}(1^+_{\bar{q}},2^-_q,3^+_g,4^+_g)$
is identical to $c_{12\x34}^{(2)}(1^+_{\bar{q}},2^-_q,3^+_g,4^+_g)$ given in
Eq.~(8.6) of ref.~\cite{Budge:2020oyl}.

\subsubsection{$\tilde c_{4\x123}^{(0)},\tilde c_{4\x123}^{(2)}$}
\begin{eqnarray}
\tilde c_{4\x123}^{(0)}(1^+_{\bar{q}},2^-_q,3^+_g,4^+_g)&=&
  -2\,(s_{14}+s_{24}+s_{34})\,
 \Bigg[
\frac{\spa2.3\,\spa2.4}{\spa1.2\,\spa3.4^3} \Bigg] \, .
\end{eqnarray}
The coefficient $\tilde c_{4\x123}^{(2)}(1^+_{\bar{q}},2^-_q,3^+_g,4^+_g)$
is identical to $c_{4\x123}^{(2)}(1^+_{\bar{q}},2^-_q,3^+_g,4^+_g)$ given in
Eq.~(8.8) of ref.~\cite{Budge:2020oyl}.

\subsubsection{$\tilde c_{3\x412}^{(0)},\tilde c_{3\x412}^{(2)}$}
\begin{eqnarray}
\tilde c_{3\x412}^{(0)}(1^+_{\bar{q}},2^-_q,3^+_g,4^+_g)&=&
   -2\,(s_{13}+s_{23}+s_{34})\,\Bigg[
   \,\frac{\spa2.3\,\spa2.4}{\spa1.2\,\spa3.4^3} \Bigg] \, .
\end{eqnarray} 
The coefficient $\tilde c_{3\x412}^{(2)}(1^+_{\bar{q}},2^-_q,3^+_g,4^+_g)$
is identical to $c_{3\x412}^{(2)}(1^+_{\bar{q}},2^-_q,3^+_g,4^+_g)$ given in
Eq.~(8.10) of ref.~\cite{Budge:2020oyl}.

\subsection{Coefficients for $A^{34}_4(\qb^+,q^-,g^-,g^+;\Higgs)$}

The coefficients that must be computed for this amplitude are shown in
the middle column of Table~\ref{table:aqgg}.

\subsubsection{$\tilde d_{3\x21\x4}$}
\begin{eqnarray}
\tilde d_{3\x21\x4}(1^+_{\bar{q}},2^-_q,3^-_g,4^+_g)&=&
   2\,m^2 \,\frac{\spab3.(1+2).4}{\spab4.(1+2).3} \,
\Bigg[\frac{\spa2.3\,\spa2.4}{\spa1.2\,\spa3.4}-\frac{\spb1.3\,\spb1.4}{\spb1.2\,\spb3.4} \Bigg] \, .
\end{eqnarray}

\subsubsection{$\tilde d_{4\x3\x21}$}
\begin{eqnarray}
\tilde d_{4\x3\x21}(1^+_{\bar{q}},2^-_q,3^-_g,4^+_g)&=&
\frac{-2}{\spab4.(1+2).3}\Bigg\{\,\frac{\spb1.3\,\spab4.(2+3).1\, s_{34}\, s_{123}^2}{\spb1.2\,\spab4.(1+2).3^2}\nonumber\\
&+& {m^2} \,\Bigg[\frac{3\,\spb1.3\,\spab3.(1+2).4\,\spab4.(2+3).1}{\spb1.2\,\spab4.(1+2).3}
+\frac{\spa2.3\,\spab2.(1+3).4}{\spa1.2} \Bigg] \Bigg\} \, .
\end{eqnarray}

\subsubsection{$\tilde c_{12\x34}^{(0)},\tilde c_{12\x34}^{(2)}$}
\begin{eqnarray}
\tilde c_{12\x34}^{(0)}(1^+_{\bar{q}},2^-_q,3^-_g,4^+_g)&=&
   8\, (s_{124}-s_{123})\, (s_{12}+s_{34}+2\,s_{13}+2\,s_{23})\,\frac{\spa2.4\,\spb1.3\,\spab3.(1+2).4}
{\spab4.(1+2).3^2 \DeltaThree(1,2,3,4)}\nonumber\\
  &+&\Big((9\,s_{13}-7\,s_{23}-s_{14}-s_{24}+4\,s_{34})\,\spa2.4\,\spb1.4 \nonumber \\
         &-&(9\,s_{14}-7\,s_{24}-s_{13}-s_{23}+4\,s_{34})\,\spa2.3\,\spb1.3\Big)\times\frac{1}{\spab4.(1+2).3^2}\nonumber\\
  &+&12\,\frac{s_{1234}\,((s_{13}+s_{23})^2-(s_{14}+s_{24})^2)\,\spab2.(3+4).1\,\spab3.(1+2).4}
     {\spab4.(1+2).3 \DeltaThree(1,2,3,4)^2}\nonumber\\
  &+&4\,
 \Big(\big\{3\,(s_{12}+s_{34})+4\,(s_{13}+s_{23}+s_{14})\big\} \, \spb1.3\,\spa2.3 \nonumber \\
   &-&\big\{3\,(s_{12}+s_{34})+4\,(s_{13}+s_{24}+s_{14})\big\} \, \spb1.4\,\spa2.4 \Big)
 \times \frac{\spab3.(1+2).4 }{\spab4.(1+2).3\,\DeltaThree(1,2,3,4)}
\nonumber \\
  &-&24\, \frac{\spb1.3\,\spa2.4\,\spab3.(1+2).4^2}{\spab4.(1+2).3\,\DeltaThree(1,2,3,4)}
  -8\,\frac{\spb1.4\,\spa2.3\,\spab3.(1+2).4}{\DeltaThree(1,2,3,4)}
  +8\,\frac{\spb1.4\,\spa2.3}{\spab4.(1+2).3} \nonumber \\
  &+&\symbrack\Bigg\{\frac{2\,\spa2.4^2\,\spb3.4\,(s_{14}+s_{24})^2}{\spa1.2\,\spab4.(1+2).3^3}
  +\frac{\spb1.3\,\spa2.4 \,(s_{14}+s_{24})\,(4\,s_{124}-2\,s_{34})}{\spab4.(1+2).3^3}\nonumber\\
  &+& \frac{2\,\spa2.3\,\spa2.4\,\spb3.4\,(s_{14}+s_{24})}{\spa1.2\,\spab4.(1+2).3^2}
-\frac{\spa2.3\,\spb1.3\,(s_{14}+s_{24}-s_{13}-s_{23})}{\spab4.(1+2).3^2}
\symbrack\Bigg\}\nonumber\\
  &-&\Bigg\{1 \leftrightarrow 2, 3 \leftrightarrow 4, \langle\,\rangle\leftrightarrow[\,] \Bigg\} \, .
\end{eqnarray}
The coefficient $\tilde c_{12\x34}^{(2)}(1^+_{\bar{q}},2^-_q,3^-_g,4^+_g)$
is identical to $c_{12\x34}^{(2)}(1^+_{\bar{q}},2^-_q,3^-_g,4^+_g)$ given in
Eq.~(9.6) of ref.~\cite{Budge:2020oyl}.

\subsubsection{$\tilde c_{4\x123}^{(0)},\tilde c_{4\x123}^{(2)}$}
\begin{eqnarray}
\tilde c_{4\x123}^{(0)}(1^+_{\bar{q}},2^-_q,3^-_g,4^+_g)&=&
 -\frac{2\,(s_{14}+s_{24}+s_{34})\,\spb1.3\,\spab4.(2+3).1\,s_{123}}{\spb1.2\,\spab4.(1+2).3^3} \, .
\end{eqnarray}
The coefficient $\tilde c_{4\x123}^{(2)}(1^+_{\bar{q}},2^-_q,3^-_g,4^+_g)$
is identical to $c_{4\x123}^{(2)}(1^+_{\bar{q}},2^-_q,3^-_g,4^+_g)$ given in
Eq.~(9.8) of ref.~\cite{Budge:2020oyl}.

\subsection{Coefficients for $A^{34}_4(\qb^+,q^-,g^+,g^-;\Higgs)$}
The coefficients for this amplitude that cannot be obtained from those for
$H^{34}_4(\qb^+,q^-,g^-,g^+)$ by performing the following operation:
$1 \leftrightarrow 2 \,, \langle\,\rangle \leftrightarrow [\,] \,$,
are listed in the right-most column of Table~\ref{table:aqgg}.
The explicit form of $\tilde c_{4\x123}$ is given here, whereas $b_{123}$
remains unaltered as compared to the fermion case.

\subsubsection{$\tilde c_{4\x123}^{(0)},\tilde c_{4\x123}^{(2)}$}
\begin{eqnarray}
\tilde c_{4\x123}^{(0)}(1^+_{\bar{q}},2^-_q,3^+_g,4^-_g)&=&
\frac{2\,\spab2.(1+3).4\,s_{123}}{\spa1.2\,\spab3.(1+2).4^2}
\Bigg\{\spa2.4
   -\,\spa3.4\frac{\spab2.(1+3).4}{\spab3.(1+2).4}\Bigg\}  \, .
\end{eqnarray}
The coefficient $\tilde c_{4\x123}^{(2)}(1^+_{\bar{q}},2^-_q,3^+_g,4^-_g)$
is identical to $c_{4\x123}^{(2)}(1^+_{\bar{q}},2^-_q,3^+_g,4^-_g)$ given in
Eq.~(10.3) of ref.~\cite{Budge:2020oyl}.

\subsection{Amplitude for $0 \to {\qb}q{\qb}q \Higgs $}

This calculation proceeds in a similar way to the calculation for a loop of fermions detailed in ref.~\cite{DelDuca:2001fn}.
The amplitude can be obtained by considering the tensor current for the scalar-mediated process $0 \to gg\Higgs$, with two off-shell gluons
(with momenta $k_1$ and $k_2$), 
\begin{equation}
{\cal T}^{\mu_1\mu_2}(k_1,k_2) = -i \delta^{c_1 c_2} \, \frac{g_s^2}{8 \pi^2} \, \Big(\frac{-\lambda}{4}\Big) \; \Big[
\tilde{F}_T(k_1,k_2) \, T_T^{\mu_1\mu_2}+ \tilde{F}_L(k_1,k_2) \, T_L^{\mu_1\mu_2}\Big] \,.
\end{equation}
The two tensor structures appearing here are,
\begin{eqnarray}
T_T^{\mu_1\mu_2} &=&k_1 \cdot k_2 \, g^{\mu_1 \mu_2}-k_1^{\mu_2} k_2^{\mu_1} \, , \\
T_L^{\mu_1\mu_2} &=&k_1^2 k_2^2 \, g^{\mu_1 \mu_2}
-k_1^2 \, k_2^{\mu_1} k_2^{\mu_2} 
-k_2^2 \, k_1^{\mu_1} k_1^{\mu_2} 
+k_1\cdot k_2 \, k_1^{\mu_1} k_2^{\mu_2} \, ,
\label{4qtensors}
\end{eqnarray}
and the form factors are given by
\begin{eqnarray}
\label{FLequation}
\label{FTequation}
\tilde{F}_T(k_1,k_2) &=& 
-\frac{1}{\Delta(k_1,k_2)} 
    \Big\{k_{12}^2 \,(B_0(k_1;m)+B_0(k_2;m)-2 B_0(k_{12};m)) -2 \,k_1\cdot k_{12} \, k_2\cdot k_{12} \, C_0(k_1,k_2;m)\nonumber \\
&+& (k_1^2-k_2^2)\,(B_0(k_1;m)-B_0(k_2;m))\Big\}-\, k_1 \cdot k_2 \, \tilde{F}_L(k_1,k_2) \, , \\
\tilde{F}_L(k_1,k_2) &=& 
-\frac{1}{\Delta(k_1,k_2)} 
    \Big\{
    \big[2-\frac{3 k_1^2 \, k_2\cdot k_{12}}{\Delta(k_1,k_2)}\big]\,(B_0(k_1;m)-B_0(k_{12};m)) \nonumber \\
&+& \big[2-\frac{3 k_2^2 \, k_1\cdot k_{12}}{\Delta(k_1,k_2)}\big]\,(B_0(k_2;m)-B_0(k_{12};m)) \nonumber \\
&-&\Big[4 m^2 +k_1^2+k_2^2 +k_{12}^2 -3 \frac{k_1^2 \,k_2^2 \, k_{12}^2}{\Delta(k_1,k_2)}\Big]\, C_0(k_1,k_2;m)-2\Big\} \, ,
\end{eqnarray}
where $k_{12}=k_1+k_2$ and $\Delta(k_1,k_2)=k_1^2 \, k_2^2 -(k_1 \cdot k_2)^2$. As expected the rational and bubble
coefficients are identical with the case for a fermion loop. 
By contracting Eq.~(\ref{4qtensors}) with currents for the quark-antiquark lines we then arrive at the result for
the amplitude.  All helicity combinations can be obtained from permutations of the single expression,
\begin{eqnarray}
A^{4q}_{4}(1^{+}_{{\qb}},2^{-}_{q},3^{+}_{{\qb}^\prime},4^{-}_{q^\prime};\Higgs)
&=&
    \Big[\frac{\spab2.(3+4).1\,\spab4.(1+2).3+\spa2.4\,\spb1.3\,(2\, p_{12}.p_{34})}{s_{12}\,s_{34}}\Big]\, \tilde{F}_T(p_{12},p_{34}) \nonumber \\
          &+& 2\, \spa2.4 \,\spb1.3\, \tilde{F}_L(p_{12},p_{34}) \,.
\end{eqnarray}

\section{Recap of inclusive and 1-jet results}
\label{sec:recap}

In this section we briefly review results for the inclusive and 1-jet cases, focussing on understanding the
pattern of results observed in the existing literature.

For all the results presented in this paper we consider the LHC operating at $\sqrt s = 14$~TeV and employ the
MMHT NLO set~\cite{Harland-Lang:2014zoa}.  In addition we use a choice of renormalization and factorization scales
appropriate for the study of Higgs+multijet events,
\begin{equation}
\mu_f = \mu_r = \frac{H_T^\prime}{2} = \frac{1}{2} \left(
 \sqrt{\mh^2 + p_{T,h}^2} + \sum_i |p_{T,i}| \right) \, ,
\end{equation}
where the sum runs over any jets (equivalently in our case, partons) present.
The mass of the top quark is $m_t = 173.3$~GeV.

\subsection{Inclusive cross section}

From Eq.~(\ref{Scalar2gluonresult2}) we can abbreviate the form of the SM and SUSY contributions to the
amplitude for inclusive Higgs production as,
\bea
{\cal M}^{SM} = {\cal H}_2^{gg} &=& C \, F_{1/2}(4m_t^2/\mh^2) \, , \\
{\cal M}^{SUSY} = {\cal A}_2^{gg}(\tilde t_1) + {\cal A}_2^{gg}(\tilde t_2)
 &=& C \left[
 \left(\frac{v}{2 m_{\tilde t_1}^2}\right) \lambda_{h \tilde t_1 \tilde t_1} F_0(4m_{\tilde t_1}^2/\mh^2)
+\left(\frac{v}{2 m_{\tilde t_2}^2}\right) \lambda_{h \tilde t_2 \tilde t_2} F_0(4m_{\tilde t_2}^2/\mh^2)
 \right] \, ,\nonumber
\eea
where $C$ is a common overall factor that is unimportant for the following argument but which can
be identified by comparison with Eq.~(\ref{Scalar2gluonresult2}).  Since we are interested in measuring deviations
from the SM result, it is useful to analyze the regions of SUSY parameter space in which these are expected to be
small and therefore hard to probe.  In order to simplify the argument we will make the simplifying assumption that
we can always work in the EFT, i.e. that $m_t, m_{\tilde t_1}, m_{\tilde t_2} \gg \mh$,  so that $F_0$ and $F_{1/2}$ can
be replaced by their asymptotic values.  This will be broadly true for the range of parameters in which we are
interested but we note that, regardless, the features we elucidate here arise even when this no longer holds.
Performing this replacement we arrive at the simple result,
\bea
{\cal M}^{SM} &=& -\frac{4C}{3} \, , \label{eq:MSM} \\
{\cal M}^{SUSY} &=& -\frac{C}{3} \left[
 \left(\frac{v}{2 m_{\tilde t_1}^2}\right) \lambda_{h \tilde t_1 \tilde t_1} 
+\left(\frac{v}{2 m_{\tilde t_2}^2}\right) \lambda_{h \tilde t_2 \tilde t_2} 
 \right] \, .
\eea
Since we are working in the limit in which the EFT is valid we can also drop the terms proportional to $\alpha_1$
and $\alpha_2$ (since they are suppressed by $m_Z^2/m_t^2$).  In that case we have the further simplification of the
SUSY amplitude,
\bea
{\cal M}^{SUSY} &=& -\frac{C}{6} \left[
 \left(\frac{m_t^2}{m_{\tilde t_1}^2}\right) \left( 2 - \frac{(\Delta m)^2}{2 m_t^2} \sin^2 2\theta \right) 
+\left(\frac{m_t^2}{m_{\tilde t_2}^2}\right) \left( 2 + \frac{(\Delta m)^2}{2 m_t^2} \sin^2 2\theta \right)
 \right] \\
&=& -\frac{C}{3} \left[
 \frac{m_t^2}{m_{\tilde t_1}^2} + \frac{m_t^2}{m_{\tilde t_2}^2}
  - \frac{1}{4} \sin^2 2\theta \frac{(\Delta m)^4}{m_{\tilde t_1}^2 m_{\tilde t_2}^2}
 \right] \, , \label{eq:MSUSY}
\eea
c.f. Eq.~(2.15) of ref.~\cite{Banfi:2018pki}.

The form of these amplitudes allows us to anticipate the situations when the SUSY contribution is very small.
Due to the mixing allowed in the top quark sector, this can occur when the SUSY amplitude itself vanishes.
It is instructive to rewrite Eq.~(\ref{eq:MSUSY}) as, 
\beq
{\cal M}^{SUSY} = \frac{C}{12 m_{\tilde{t}_1}^2 m_{\tilde{t}_2}^2} \left[
 \sin^2 2\theta (\Delta m)^4 -4 m_t^2 (\Delta m)^2 - 8 m_t^2 m_{\tilde{t}_1}^2
 \right] \, ,
\eeq
so that the dependence on $m_{\tilde{t}_2}$ in the numerator has been eliminated.
From this it is clear for which values of $m_{\tilde{t}_1}$, $\Delta m$
and $\theta$ the amplitude vanishes, so that the SUSY result is very close to the SM one.
Solving for ${\cal M}^{SUSY} = 0$ we find this occurs when ($\theta > 0$),
\beq
\Delta m = m_t \times \sqrt{2} \times \sqrt{\frac{ 1 +
 \sqrt{1 + 2 \sin^2 2\theta \, m_{\tilde{t}_1}^2/m_t^2}}
 {\sin^2 2\theta }} \, .
\label{eq:deltazero1}
\eeq

The coincidence of the SM and SUSY cross sections discussed above arises from a vanishing
of the SUSY amplitude.  In addition, there can be a further coincidence when the effect of
interference between the SUSY and SM contributions cancels the contribution from the SUSY
amplitude squared.  In other words we must have an alternative solution when,
\bea
&& \left({\cal M}^{SM} + {\cal M}^{SUSY} \right)^2 - \left({\cal M}^{SM}\right)^2 = 0 \nonumber \\
&\implies& {\cal M}^{SUSY} + 2{\cal M}^{SM} = 0 \, ,
\eea

Using the results in Eqs.~(\ref{eq:MSM}) and~(\ref{eq:MSUSY}) we have,
\beq
{\cal M}^{SUSY} + 2{\cal M}^{SM} =
 -\frac{C}{3} \left[
 \frac{m_t^2}{m_{\tilde t_1}^2} + \frac{m_t^2}{m_{\tilde t_2}^2}
  - \frac{1}{4} \sin^2 2\theta \frac{(\Delta m)^4}{m_{\tilde t_1}^2 m_{\tilde t_2}^2}
  + 8 \right] \, .
\eeq
Manipulating as above, we find that this vanishes when,
\beq
\sin^2 2\theta  (\Delta m)^4 - 4 (m_t^2 + 8 m_{\tilde{t}_1}^2) (\Delta m)^2
  - 8 m_{\tilde{t}_1}^2 (m_t^2 + 4 m_{\tilde{t}_1}^2) = 0 \, .
\eeq
This equation has no solutions for small $\Delta m$.  On the other hand, for
large $\Delta m$ the SUSY and SM cross sections are identical when,
\beq
\Delta m \approx  2 \frac{\sqrt{m_t^2 + 8 m_{\tilde{t}_1}^2}}{\sin 2\theta} \, .
\label{eq:vanishxsec}
\eeq
Again, this solution relies on the existence of a non-zero mixing ($\theta > 0$) between the squarks.

\begin{figure}[t]
  \begin{center}
  \includegraphics[width=1.0\textwidth]{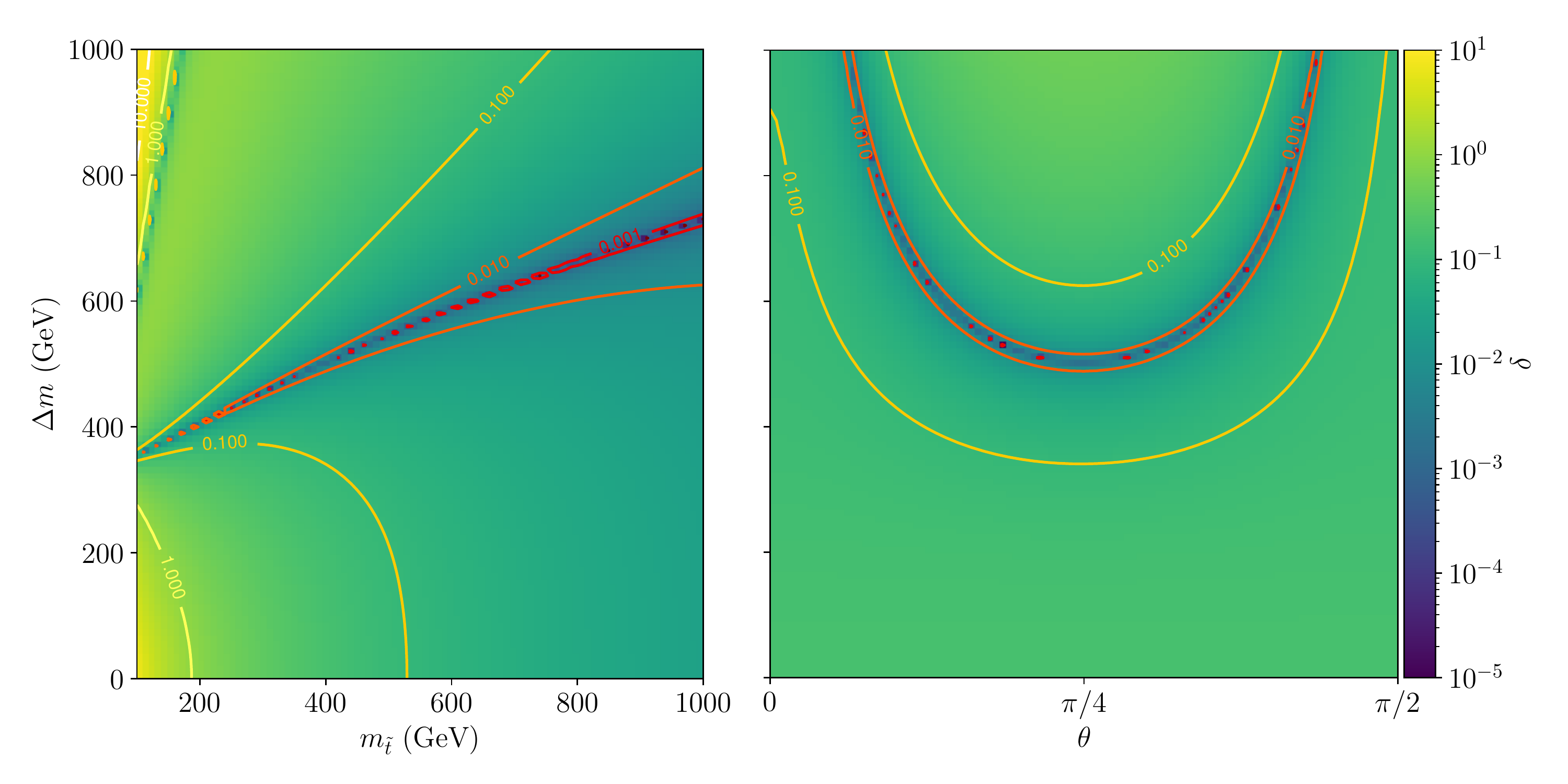}
  \caption{The deviation of the inclusive Higgs cross section from the SM case, measured by $\delta$ defined
  in Eq.~(\ref{eq:delta}). The left panel shows $\delta$ as a function of $m_{\tilde t_1}$ and $\Delta m$ with top squarks mixing in a
  maximal fashion ($\theta=\frac{\pi}{4}$). In the right panel $\delta$ is shown as a function of $\theta$ and $\Delta m$, $m_{\tilde t_1}=400$ GeV.   
  In both panels $\tan\beta=10$.}
  \label{figure:0jet_mdeltaM}
  \end{center}
\end{figure}
We now turn to a numerical study, measuring the deviation between the 
SUSY and SM cases by the quantity $\delta$ defined as~\cite{Banfi:2018pki},
\beq
\delta = \left| \frac{\sigma^{SM}-\sigma^{SUSY}}{\sigma^{SM}} \right| \,.
\label{eq:delta}
\eeq
As discussed above, the case of non-zero mixing is most interesting; in the absence of any mixing the SUSY
contribution is simply additive.   For this reason we focus on the case of maximal mixing ($\theta=\pi/4$)
to illustrate the pattern of behavior.  Results for the case $\tan\beta=10$, and as a function of the parameters
$m_{\tilde t_1}$ and $\Delta m$, are shown in Fig.~\ref{figure:0jet_mdeltaM} (left). 
This figure demonstrates the regions of vanishing $\delta$ anticipated above.  First, the vanishing of the 
SUSY amplitude occurs, in the maximal-mixing case, for values of $\Delta m$ given by,
\beq
\Delta m \approx m_t \sqrt{2 \left(1 + \sqrt{2} \, m_{\tilde{t}_1}/m_t \right)}.
\eeq
This corresponds to the dark blue stripe across the middle of the plot in Fig.~\ref{figure:0jet_mdeltaM} (left),
already observed in Ref.~\cite{Banfi:2018pki}.  The cancellation at the level of the cross section, i.e. as
expected from Eq.~(\ref{eq:vanishxsec}), corresponds to the lighter-blue line in the upper-left corner of
the plot.  In the region shown, $m_{\tilde{t}_1} \approx m_t$, it is approximately given by
$\Delta m \approx 6 m_{\tilde{t}_1}$.

Note that, although we have used asymptotic results for the amplitudes to derive the presence and locations of
the features above, these are clearly sufficient to capture the dominant effects.  At smaller values of
$m_{\tilde{t}_1}$, and to some extent $\Delta m$, the precise contours of vanishing $\delta$ vary slightly but
are still present.

Although the maximal-mixing case is of highest interest here, as an indication of the effect of a smaller amount
of mixing, Fig.~\ref{figure:0jet_mdeltaM} (right) shows similar contours as a function of $\theta$ and $\Delta m$,
for fixed $m_{\tilde t_1}=400$~GeV.  Again the region of vanishing $\delta$ that is clearly visible in the figure
is easily understood from Eq.~(\ref{eq:deltazero1}).

\subsection{1-jet cross section}

\begin{figure}[t]
\begin{center}
\includegraphics[angle=270,width=0.7\textwidth]{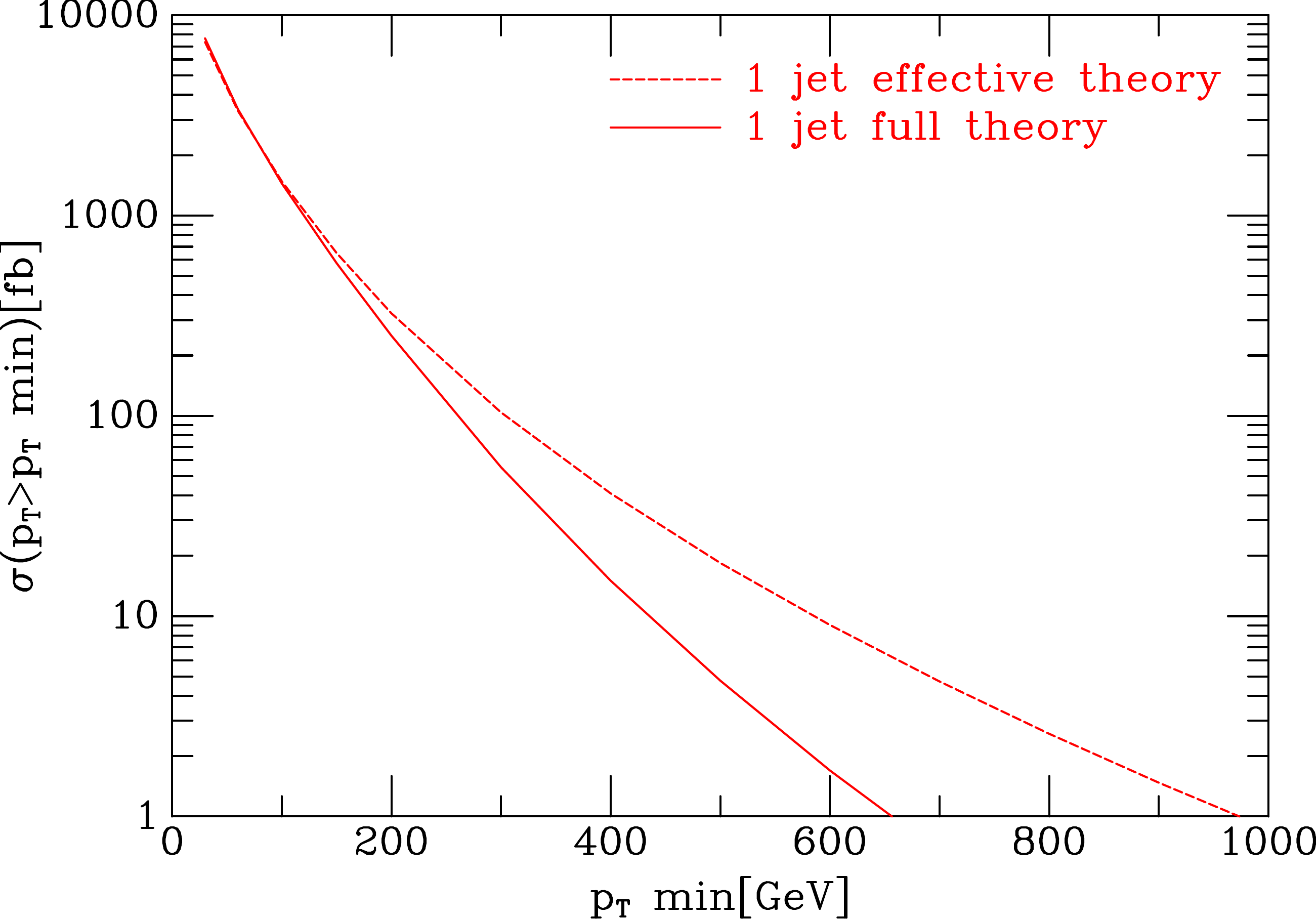}
\caption{Lowest order predictions for H+1 jet in the SM model, computed in the EFT (dashed) and
in the full theory (solid).}
\label{figure:H1jet}
\end{center}
\end{figure}
The case when the Higgs boson is produced in association with a jet has also been discussed extensively in
the literature~\cite{Brein:2007da,Grojean:2013nya,Banfi:2018pki}.  As explained in Ref.~\cite{Banfi:2018pki},
the pattern of deviations from the SM is very similar to the inclusive case for low-$p_T$ jets, but begins
to differ as the jet (or equivalently, Higgs boson) is produced at large transverse momentum.  A simple way
to understand this behaviour is in terms of the ability of the jet to resolve the loop of coloured particles,
which is demonstrated for the SM case in Fig.~\ref{figure:H1jet}.  To compute the cross sections shown in this figure we have
defined jets according to a minimum $p_T$, also satisfying the requirement $|y({\rm jet})|<2.4$.
As discussed earlier, the calculations in the full theory and in the EFT only begin to differ around
$p_T \sim 200$~GeV, i.e. when the jet is sufficiently energetic to resolve the top-quark loop.  The same applies,
of course, for the SUSY amplitude, except that the appropriate scales are now set by $m_{\tilde t_1}$ and
$\Delta m$.  In the low-$p_T$ region both the SM and  SUSY contributions to the amplitude can be computed
in the effective theory and thus the pattern of deviations from the SM must be very similar to the inclusive
case.   Indeed, since the equality of the SM and SUSY cross sections is tied to the structure of the SUSY amplitude,
the $p_T$ of the jet should be of order $m_{\tilde t_1}$ in order to probe the dependence on the SUSY parameters more
effectively.

\begin{figure}[t]
  \begin{center}
  \includegraphics[width=1.0\textwidth]{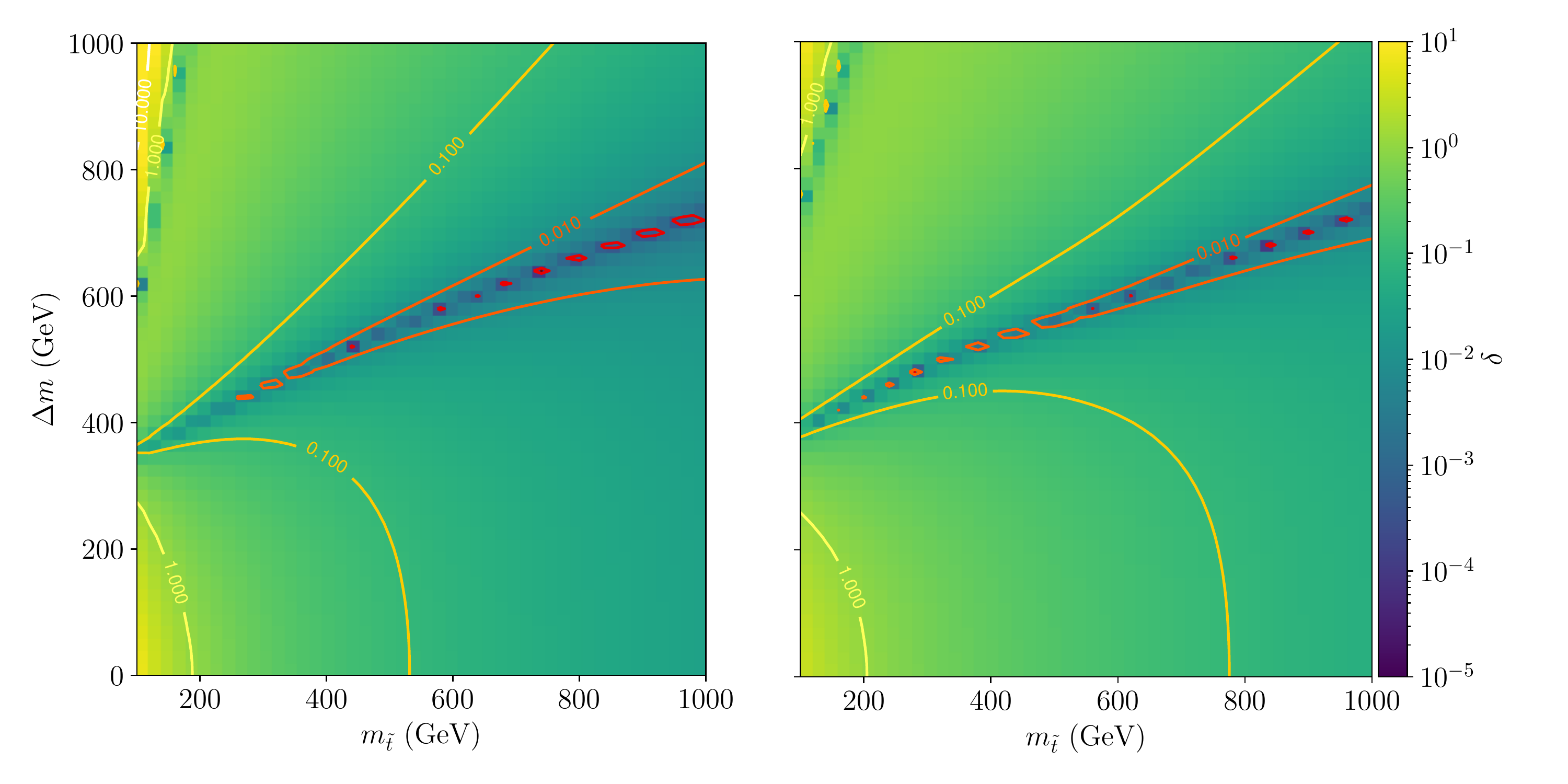}
  \caption{The deviation of the Higgs+1 jet cross section from the SM case, measured by $\delta$ defined
  in Eq.~(\ref{eq:delta}), as a function of $m_{\tilde t_1}$ and $\Delta m$.  Top squarks mix in a
  maximal fashion ($\theta=\frac{\pi}{4}$) and $\tan\beta=10$.
  Results are shown for two choices of jet $p_T$: 30~GeV (left) and 600~GeV (right).}
  \label{figure:1jet_mdeltam}
  \end{center}
\end{figure}

This argument has been discussed in great detail in Ref.~\cite{Banfi:2018pki}, from which we reproduce one
set of results in order to illustrate this point. Fig.~\ref{figure:1jet_mdeltam} shows the
dependence of $\delta$ on $m_{\tilde{t}}$ and $\Delta m$ for the 1-jet process, for two choices of minimum
jet (Higgs) $p_T$.   For the case of a low jet $p_T$ cut, at $30$~GeV (Fig.~\ref{figure:1jet_mdeltam}, left panel),
the deviations from the SM case are essentially identical to the 0-jet case (c.f. Fig.~\ref{figure:0jet_mdeltaM}, left).
For a much higher cut, at 600~GeV,
the jet is able to resolve at least the top quark loop and also the top squark, for values of $m_{\tilde t_1}$
up to a similar scale.  This breaks the similarity with the 0-jet case, leading to larger deviations from the SM.
This is most visible in the contours of constant $\delta$, that are now more constraining for the 1-jet case.
This confirms the results presented in Ref.~\cite{Banfi:2018pki}.

The region where the lightest top squark is degenerate with the top quark,
$m_{\tilde t_1} \approx m_t$, has received considerable interest due to the
difficulty of directly detecting a signal from such a model (see, for instance,
Ref.~\cite{Cohen:2019ycc} and references therein).  Although in general this
scenario reults in large corrections to the Higgs boson rate,
the blue lines in Fig.~\ref{figure:0jet_mdeltaM}
(left) indicate the two regions where deviations are small and an indirect
search via the inclusive Higgs cross section is similarly insensitive.
However, applying a sufficiently high cut on the transverse momentum
of the Higgs boson modifies both of these regions
(Fig.~\ref{figure:1jet_mdeltam}, right) and such scenarios could be
excluded by comparing the 0- and 1-jet rates.

\section{Results for the 2-jet process}
\label{sec:pheno}

For the 2-jet case we must supplement the jet $p_T$ and rapidity threshold by a proper jet clustering algorithm.
For this we choose the anti-$k_T$ algorithm with a jet resolution parameter $R=0.5$.

\begin{figure}[t]
\begin{center}
\includegraphics[angle=270,width=0.7\textwidth]{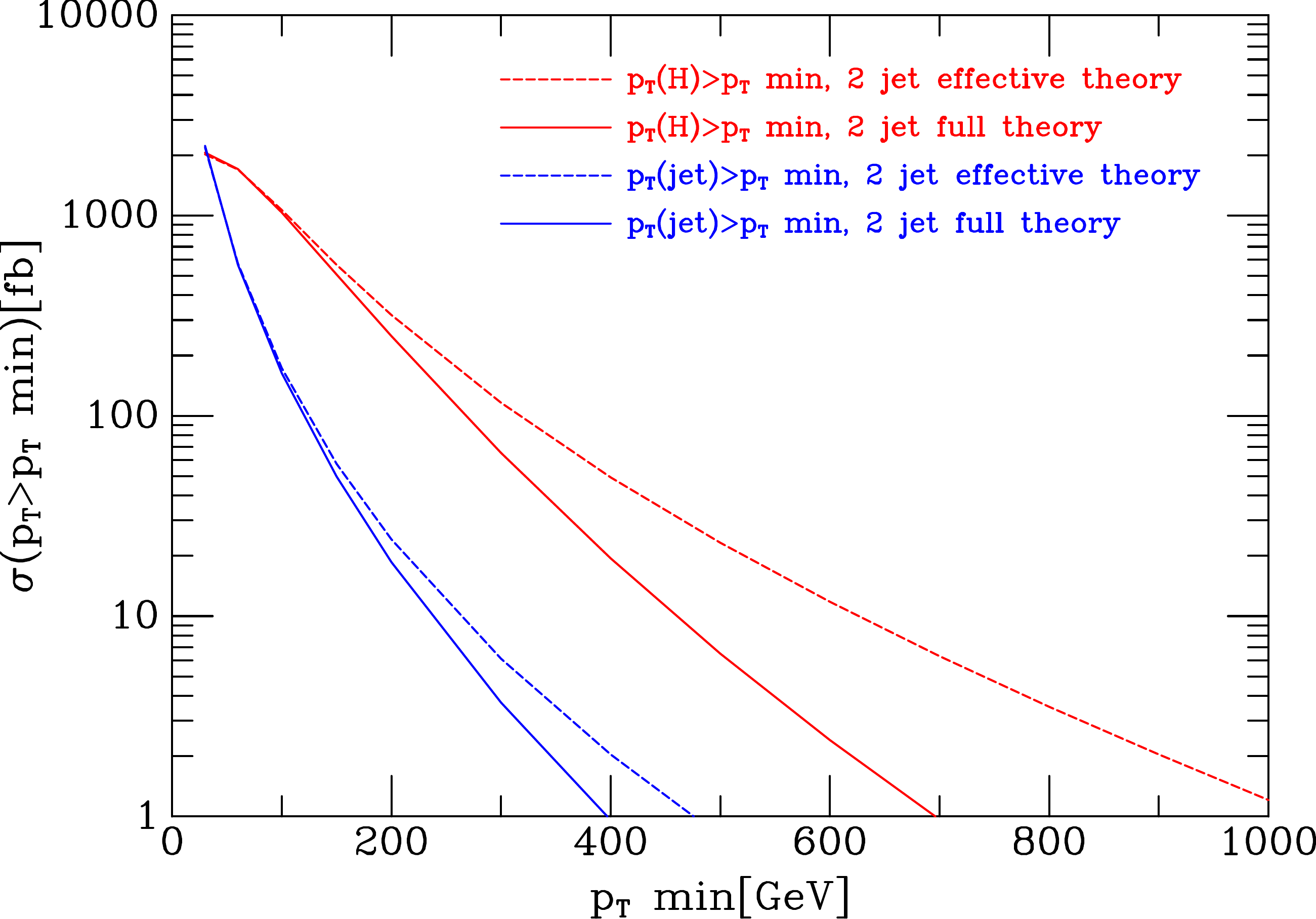}
\caption{Rates for H+2 jet production in the SM, as a function of a minimum $p_T$, computed
in the full theory (solid) and EFT (dashed).  Jets are either subject to this minimum $p_T$ themselves
(blue), or they are only required to satisfy a $30$~GeV cut and the minimum $p_T$ cut is applied to the
Higgs boson $p_T$ (red).}
\label{figure:H2jet}
\end{center}
\end{figure}

As we have already discussed, differences between the pattern of cross-section deviations are intimately
connected to the breakdown of the EFT approach to describing these processes.  We therefore first examine
this for the 2-jet case in the SM, with the results shown in Fig.~\ref{figure:H2jet}.
As the jet $p_T$ cut is increased, the difference between the EFT and the full theory is not as pronounced
in the 2-jet case as in the 1-jet process (comparing blue curves in Fig.~\ref{figure:H2jet} with
Fig.~\ref{figure:H1jet}).  As explained in Ref.~\cite{Greiner:2016awe}, which explored the limitations of the EFT
by studying Higgs+1, 2 and 3 jet processes, this is because the breakdown of the EFT is
controlled by the $p_T$ of the single hardest particle in the process.  Requiring two very hard jets only
serves to decrease the rate without providing an additional probe of the loop-induced Higgs coupling. Therefore, in order to
drive the EFT breakdown more efficiently, and thus observe a different pattern of dependence
on the SUSY parameters, we should employ a cut that requires a single hard particle.  Therefore we choose
to cluster jets with the usual cut, $p_T({\rm jet}) > 30$~GeV, and then make a cut on the $p_T$ of the Higgs boson.
In this case the difference between the full theory and the EFT is very similar in the 1- and 2-jet cases
(comparing red curves in Fig.~\ref{figure:H2jet} with Fig.~\ref{figure:H1jet}).

\begin{figure}[t]
\begin{center}
\includegraphics[angle=270,width=0.7\textwidth]{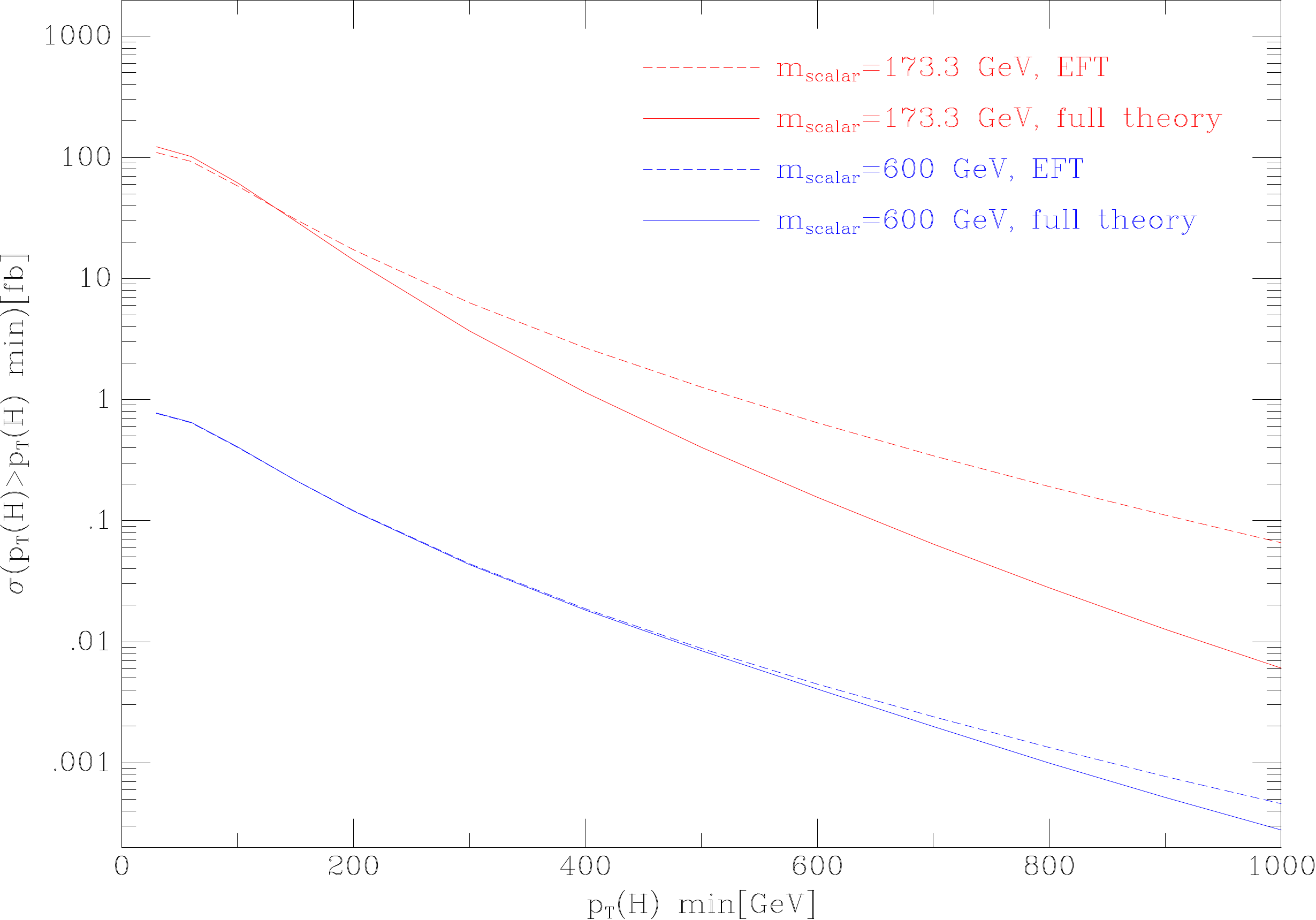}
\caption{Rates for H+2 jet production through a scalar loop, as a function of a minimum $p_T$ applied
to the Higgs boson, computed in the full theory (solid) and EFT (dashed).
The scalar mediator mass is either $173.3$~GeV (red) or $600$~GeV (blue).}
\label{figure:H2jetscalar}
\end{center}
\end{figure}
\begin{figure}[t]
\begin{center}
\includegraphics[angle=270,width=0.7\textwidth]{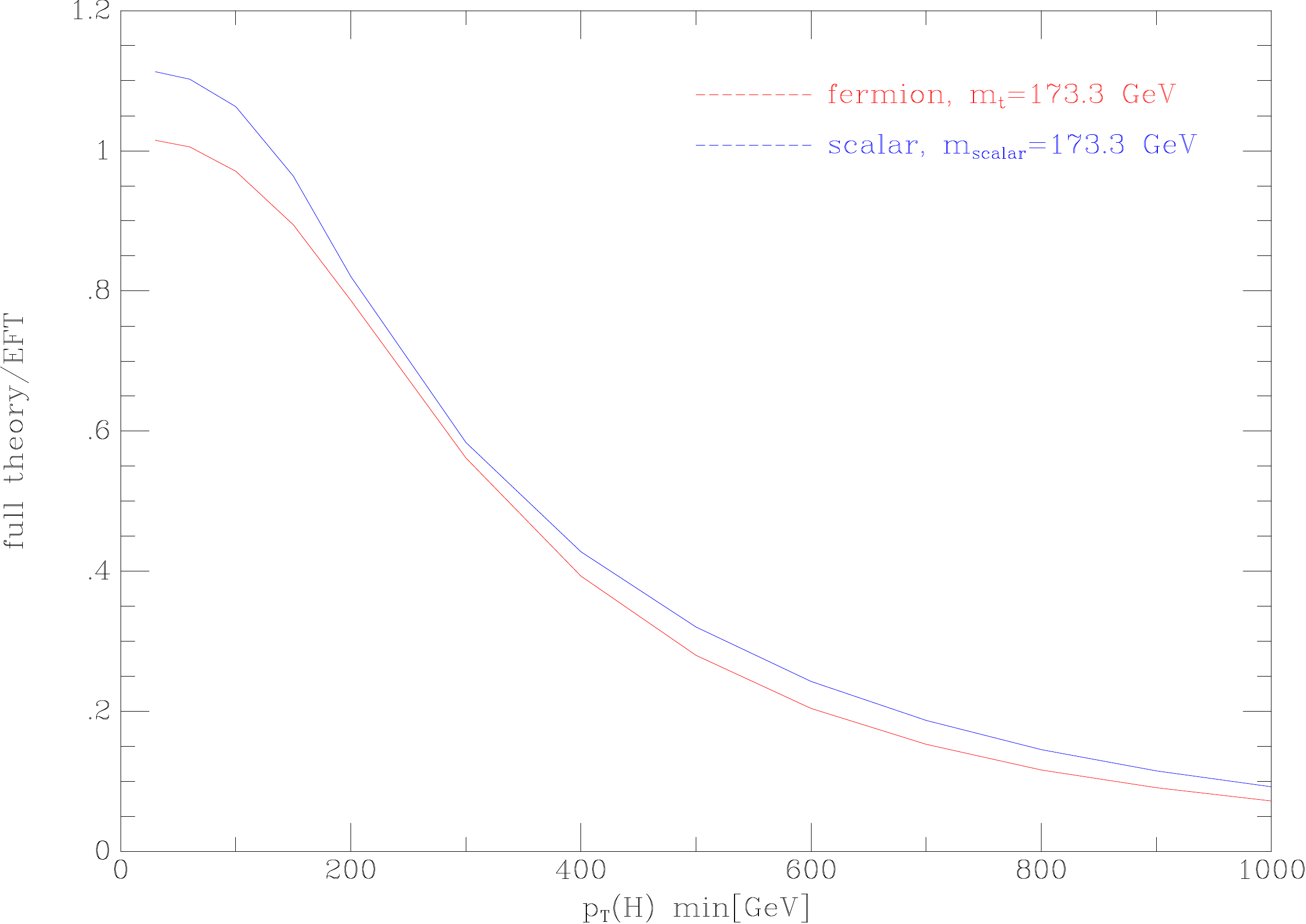}
\caption{The ratio of the cross-section in the full theory to that in the effective theory, for
H+2 jet production through a fermion (red) and scalar (blue) loop, as a function of a minimum $p_T$ applied
to the Higgs boson. The mass of the mediator is set to $173.3$~GeV in both cases.}
\label{figure:H2jetratio}
\end{center}
\end{figure}

We now use this same cut to explore the breakdown of the EFT for the case of a scalar particle
in the loop, with results shown in Fig.~\ref{figure:H2jetscalar}.  For this study we consider only the effect
of a single scalar particle in the loop and no top quark, with the amplitudes for the scalar-mediated
EFT implemented according to the discussion in appendix~\ref{sec:largemass}.
For a direct comparison with the fermion case we show
results for $m_{\rm scalar} = m_t = 173.3$~GeV, and also for a much higher mass, $m_{\rm scalar} =600$~GeV.
As expected, for each case the breakdown of the EFT occurs for $p_T(H, {\rm min}) \sim m_{\rm scalar}$.  As illustrated in
Fig.~\ref{figure:F}, for the inclusive process, the behaviour of the fermionic and scalar amplitudes in the
vicinity of the 2-particle threshold differs.  However, for the 2-jet case, any such difference is not
reflected at the level of the cross-section, as shown in Fig.~\ref{figure:H2jetratio}.  Although the effective
theory appears to work a little better for high $p_T({\rm min})$ in the scalar case, overall the two curves are very
similar.  In the limit of small $p_T({\rm min})$ the result in the full theory is actually larger
than the one computed in the EFT, in both cases.  This is expected from the inclusive calculation,
which these results should
resemble as $p_T({\rm min}) \to 0$, where the ratios can be computed from Eqs.~(\ref{Fzero})
and~(\ref{Fhalf}) (c.f. also Fig.~\ref{figure:F}),
\bea
\sigma^{\rm full}(gg \to H)/\sigma^{\rm EFT}(gg \to H) |_{\rm fermion}
 &=& \left[F_{1/2}(4m_t^2/\mh^2)/(-4/3)\right]^2 = 1.065 \,,\nonumber \\
\sigma^{\rm full}(gg \to H)/\sigma^{\rm EFT}(gg \to H) |_{\rm scalar}
 &=& \left[F_{0}(4m_t^2/\mh^2)/(-1/3)\right]^2 = 1.157 \,.
\eea
Although the two curves in Fig.~\ref{figure:H2jetratio} never reach these values, due to the presence
of the additional jets, they do reflect this underlying difference in the quality of the EFT.  
 
\begin{figure}[t]
  \begin{center}
  \includegraphics[width=0.6\textwidth,angle=90,trim=120 120 120 80,clip]{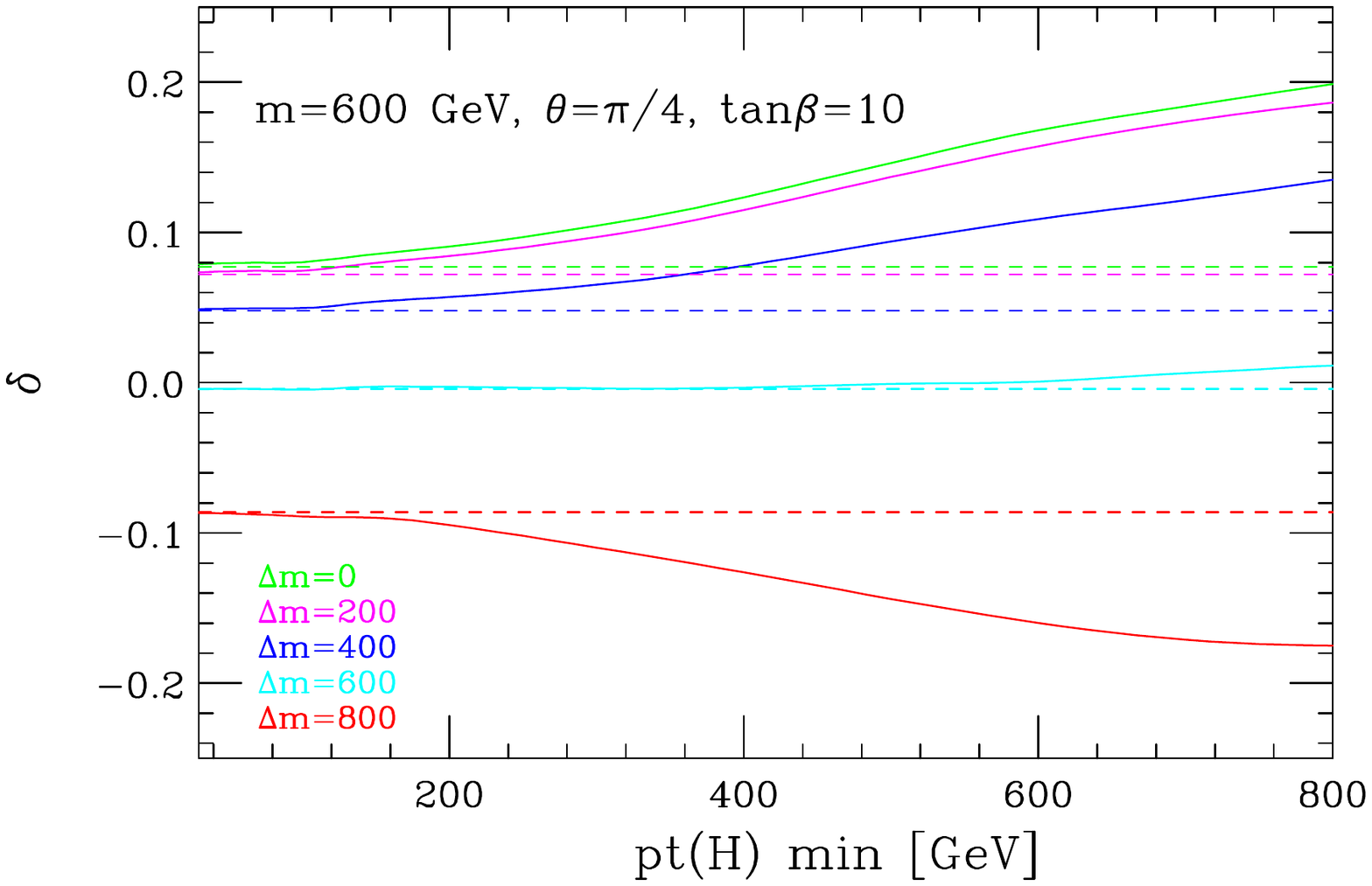}
  \caption{2-jet calculation of $\delta$ (solid lines), as a function of the cut on the Higgs boson
           $p_T$,  for the parameters $m_{\tilde t_1}=600$~GeV,
           $\theta=\frac{\pi}{4}$ and $\tan\beta=10$.  The corresponding inclusive result is shown as a dashed line.}
    \label{fig:2jetpth}
  \end{center}
\end{figure}
\begin{figure}[t]
  \begin{center}
  \includegraphics[width=1.0\textwidth]{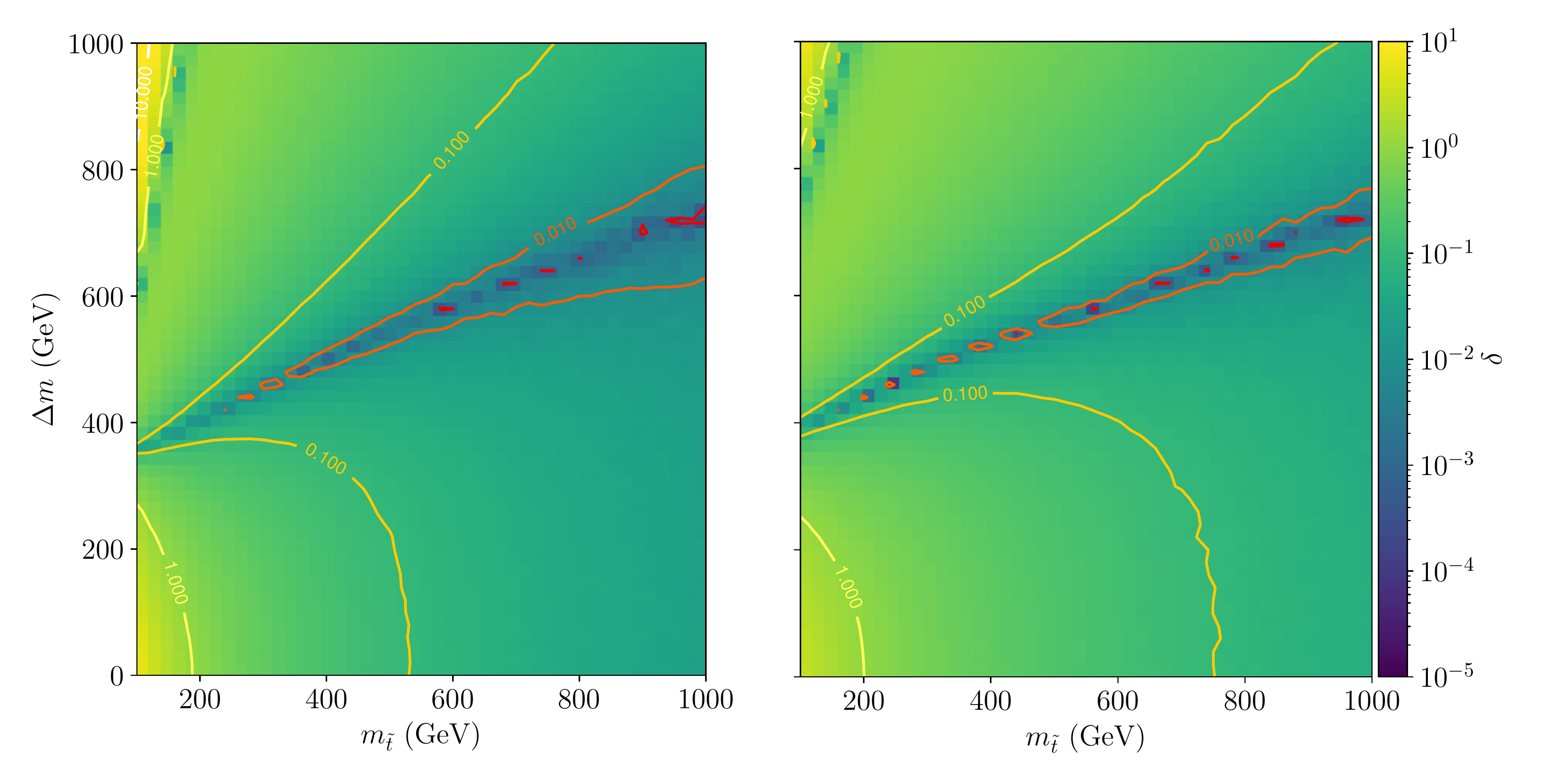}
  \caption{The deviation of the Higgs+2 jet cross section from the SM case, measured by $\delta$ defined
  in Eq.~(\ref{eq:delta}), as a function of $m_{\tilde t_1}$ and $\Delta m$.  Top squarks mix in a
  maximal fashion ($\theta=\frac{\pi}{4}$) and $\tan\beta=10$.
  Results are shown in the cases of no additional cut (left) and $p_T(H) > 600$~GeV (right).}
  \label{figure:2jet_mdeltam_John}
  \end{center}
\end{figure}

To examine the sensitivity of the 2-jet process to the SUSY parameters we again focus on the maximal-mixing case.
We first assess the dependence on the minimum Higgs $p_T$ cut that is applied, for the case of
$m_{\tilde t_1}=600$~GeV.  The results are shown in Fig.~\ref{fig:2jetpth}.  Note that we have covered a range of
$p_T$ that we expect to be accessible at the LHC -- the cross-section above $800$~GeV is less than $0.5$fb, so around
1500 such Higgs events in the full HL-LHC dataset, 3ab$^{-1}$.  As indicated in the figure, the deviations are bigger in the 2-jet
case than for the inclusive cross section.  However, the results are almost identical to the 1-jet case, c.f.
Fig.~2 of Ref.~\cite{Banfi:2018pki}.  This is further reflected in the expected deviations from the SM
shown in Fig.~\ref{figure:2jet_mdeltam_John} -- a different pattern from the 0-jet process, but almost identical
to the 1-jet results shown in Fig.~\ref{figure:1jet_mdeltam}.

\subsection{Discussion}

In evaluating the discriminating power in the 0-, 1- and 2-jet cases above, we have focussed only on the deviations
between the rates in the SM and SUSY cases.  However, in order to observe such a difference, one must
also take into account the number of events that could actually be produced in each case.  Reading off the
cross-sections from figures~\ref{figure:H1jet} and~\ref{figure:H2jet} for $p_T({\rm jet})> 30~{\rm GeV}$,
and using the (similarly leading-order) result for the inclusive cross-section we have,
\bea
\sigma(gg \to H) &=& 16240~{\rm fb} \,, \nonumber \\
\sigma(gg \to H+1~{\rm jet}) &=& 7640~{\rm fb} \,,\qquad \nonumber \\
\sigma(gg \to H+2~{\rm jets}) &=& 2230~{\rm fb} \,,\qquad \nonumber \\
\sigma(gg \to H+1~{\rm jet}) &=& 2~{\rm fb} \qquad (p_T(H)> 600~{\rm GeV}) \,, \nonumber \\
\sigma(gg \to H+2~{\rm jets}) &=& 2~{\rm fb} \qquad (p_T(H)> 600~{\rm GeV}) \,.
\eea
Since the number of expected events with a highly-boosted Higgs boson is so small, for both the 1-
and 2-jet cases, it appears unlikely that the extra discriminating power could ever come into play.  Even
accounting for the fact that it may be possible to identify the $H \to b\bar b$ decay of the Higgs boson
in the boosted case -- essentially impossible for the bulk of Higgs boson events that occur at low $p_T$ -- the expected
event sample would be at least two orders of magnitude smaller than any of the non-boosted ones.
In such a case the improved discrimination between the SM and SUSY scenarios would never overcome the loss
in statistical power.

Of course, in making this argument we have neglected the role of systematic uncertainties.
On the experimental side, the boosted configuration of the Higgs boson means that its decay products are more energetic
and thus may be measured with smaller uncertainties.  On the other hand, the precision of the SM theoretical prediction
with which the data must be compared is much reduced: going from current percent-level uncertainties in an expansion up to
N$^3$LO at the inclusive level~\cite{Anastasiou:2016cez,Mistlberger:2018etf} to $10\%$ level uncertainties from
NNLO in the boosted case~\cite{Chen:2014gva,Chen:2016zka,Bizon:2018foh,Boughezal:2015dra,Campbell:2019gmd}.  While it
is clear that Higgs+multijet events offer a different handle on signals of new physics, a proper
accounting of both statistical and systematic uncertainties is essential to fully understand their value.

\section{Conclusions}
\label{sec:conclusions}
In this paper we have provided an analytic calculation of all amplitudes representing
the scattering of a Higgs boson and four partons, mediated by a loop of color-triplet
scalar particles.\footnote{A computer-readable form of all the integral coefficients presented
in section~\ref{sec:coeffs} -- which, together with the results in Ref.~\cite{Budge:2020oyl}, is sufficient to
reproduce these amplitudes -- is provided at {\tt https://mcfm.fnal.gov/scalarcoeffs.tar.gz}.}
By combining this with a previous calculation of the corresponding
amplitudes in which the mediator particle is a fermion~\cite{Budge:2020oyl} we are able
to describe modifications to the SM production of a Higgs boson in association with two jets
in theories containing such scalar extensions.  As an example we have analyzed the
specific case of the MSSM, which contains two relevant scalar particles,
$\tilde t_1$ and $\tilde t_2$.  Sensitivity to this scenario has previously been
considered extensively in the literature, for both inclusive production and the case
of Higgs production in association with one jet.  Our study is the first time such an
analysis has been performed for the case of two jets.

The results of our calculation show that, although a 2-jet analysis offers improved
sensitivity compared to an inclusive analysis, it does not provide an additional benefit
over the 1-jet case.  This can be understood by noting that, in order to probe the nature
of the loop process most effectively, the 2-jet analysis should demand only a single hard
particle:  either a jet, or the Higgs boson itself.  In either case the cross section is
dominated by configurations in which the Higgs boson recoils against a single hard jet,
with the second jet relatively soft.  Such configurations are therefore 1-jet-like,
with the emission of the second jet well described by the QCD properties of soft and
collinear factorization.  Moreover, the loss in statistical power that ensues from selecting
a sample of events in such a configuration cannot overcome the relatively-small improvement
in sensitivity.  The same conclusion applies to studies of the 1-jet rate at high transverse
momentum.

Finally, we note that if a deviation from the SM prediction were observed in a sample
of events containing Higgs bosons produced at high transverse momentum, it would be
essential to have precision theoretical predictions for such configurations in a variety
of beyond-the-SM scenarios.  The amplitudes presented in this paper are an ingredient
in a next-to-leading order calculation of the Higgs+jet process in theories containing
colour-triplet scalars.

\section*{Acknowledgements}
We thank Giuseppe de Laurentis for his input and collaboration in the early
stages of this project. 
The numerical calculations reported in this paper were performed using the Wilson High-Performance
Computing Facility at Fermilab and the Vikram-100 High Performance Computing Cluster at Physical Research Laboratory. 
This document was prepared using the resources of
the Fermi National Accelerator Laboratory (Fermilab), a
U.S. Department of Energy, Office of Science, HEP User
Facility. Fermilab is managed by Fermi Research Alliance, LLC (FRA),
acting under Contract No.\ DE-AC02-07CH11359.

\appendix

\section{Numerical value of coefficients at a given phase-space point}
\label{sec:PScheck}

Tables~\ref{CoeffValues} and~\ref{aqggtable} contain numerical results for the integral
coefficients for the $ggggh$ and $\bar q q gg h$ amplitudes respectively,
at the phase-space point ($p = (E,p_x,p_y,p_z)$)
\beqn \label{kinpoint_paper}
p_1      &=&(-15\kappa,-10\kappa, +11\kappa,  +2\kappa) \, ,\nonumber \\
p_2      &=&(-9\kappa,  +8\kappa,  +1\kappa,  -4\kappa) \, ,\nonumber \\
p_3      &=&(-21\kappa, +4\kappa, -13\kappa, +16\kappa) \, ,\nonumber \\
p_4      &=&(-7\kappa,  +2\kappa,  -6\kappa,  +3\kappa) \, ,\nonumber \\
p_\Higgs &=&(+52\kappa, -4\kappa,  +7\kappa, -17\kappa) \, ,
\eeqn
with $\kappa=1/\sqrt{94}$ and $p_\Higgs=-p_1-p_2-p_3-p_4$.
This fixes $s_{1234}=25$, $\mh=5$ and we further choose $m=1.5$.

Due to the correspondence of results between the scalar and fermionic loop cases,
many of these values have already been reported in ref.~\cite{Budge:2020oyl}.
Tables~\ref{CoeffValues} and~\ref{aqggtable}
therefore contain only the coefficients that differ from the fermionic case. 
\begin{table}
\begin{center}
\begin{tabular}{|l|l|l|l|l|}
\hline
Helicities & Coefficient        & Real Part     & Imaginary Part & Absolute Value \\
\hline
+\,+\,+\,+ &  $\tilde{d}_{1\x2\x34}$  &    -0.9840613828 &    -0.5144323508 &     1.1104131883 \\
           &  $\tilde{d}_{1\x23\x4}$  &    -3.3548957407 &    -4.8432206981 &     5.8916985803 \\
           &  $\tilde{d}_{1\x2\x3}$   &    -6.7445910748 &   -15.4663942318 &    16.8730216411 \\
           &  $\tilde{c}_{1\x234}$    &   -10.6368762164 &   -31.6829840771 &    33.4208709592 \\
\hline
+\,+\,+\,-- &  $\tilde{d}_{1\x2\x34}$  &    23.4451295603 &    18.5996441921 &    29.9269254046 \\
           &  $\tilde{d}_{1\x4\x32}$  &    20.5071688388 &    27.4451393815 &    34.2604677355 \\
           &  $\tilde{d}_{2\x1\x43}$  &    -4.9009936782 &    42.1225176136 &    42.4066767047 \\
           &  $\tilde{d}_{2\x34\x1}$  &   -44.3845463184 &   -38.3339964812 &    58.6471076705 \\
           &  $\tilde{d}_{4\x3\x21}$  &    -7.1203811993 &     0.6886216537 &     7.1536024635 \\
           &  $\tilde{d}_{1\x23\x4}$  &    -1.8005835535 &     1.5351129014 &     2.3661514646 \\
           &  $\tilde{d}_{2\x3\x4}$   &     0.8206155641 &     1.4735210192 &     1.6866161680 \\
           &  $\tilde{d}_{1\x2\x3}$   &   -19.2397847846 &    -1.4762925832 &    19.2963405429 \\
           &  $\tilde{d}_{3\x4\x1}$   &    -0.3316788675 &     1.6114692592 &     1.6452489309 \\
           &  $\tilde{c}_{4\x123}$    &   -11.0616538761 &    -1.7916339105 &    11.2058082504 \\
           &  $\tilde{c}_{1\x234}$    &    18.9646702722 &    24.4510167733 &    30.9436736633 \\
           &  $\tilde{c}_{2\x341}$    &    -8.9934514290 &    11.1934355822 &    14.3588010899 \\
           &  $\tilde{c}_{12\x34}$    &    -3.7461389306 &    21.0493483972 &    21.3800988032 \\
\hline
+\,--\,+\,-- &  $\tilde{d}_{4\x3\x21}$  &    -6.9368235764 &   -13.4220769362 &    15.1086621053 \\
           &  $\tilde{d}_{1\x23\x4}$  &    -5.4005161311 &     3.8281939917 &     6.6197162870 \\
           &  $\tilde{d}_{1\x2\x3}$   &   -21.0997803781 &   -62.3608308275 &    65.8336840341 \\
           &  $\tilde{c}_{12\x34}$    &   -39.7403340718 &    22.2104113517 &    45.5257786814 \\
           &  $\tilde{c}_{1\x234}$    &     3.9682125956 &    13.4813791531 &    14.0532663489 \\
\hline
+\,+\,--\,-- &  $\tilde{d}_{1\x2\x34}$  &    -0.0267530609 &    -1.1100908623 &     1.1104131883 \\
           &  $\tilde{d}_{1\x4\x32}$  &    22.6518970482 &  -458.1248398611 &   458.6845074097 \\
           &  $\tilde{d}_{2\x34\x1}$  &    64.2316548189 &   -59.0233562841 &    87.2322306708 \\
           &  $\tilde{d}_{1\x23\x4}$  &    -5.7450785528 &     3.2885735727 &     6.6197162870 \\
           &  $\tilde{d}_{1\x2\x3}$   &   -10.8954346530 &   -12.8836471165 &    16.8730216411 \\
           &  $\tilde{c}_{23\x41}$    &  1075.3186068541 &   747.6290891424 &  1309.6791061854 \\
           &  $\tilde{c}_{1\x234}$    &    36.8856220760 &  -309.1172377677 &   311.3101601314 \\
\hline
\end{tabular}
\caption{Numerical values of coefficients of the $ggggh$ process not already reported in ref.~\cite{Budge:2020oyl} at kinematic point, \ref{kinpoint_paper}}
\label{CoeffValues}
\end{center}
\end{table}
  \begin{table}
 \begin{center}
 \begin{tabular}{|l|l|l|l|l|}
 \hline
 Helicities & Coefficient & Real Part & Imaginary Part & Absolute Value \\
\hline
+\,--\,+\,+ &$d_{3\x21\x4}$&    370.4335392027&   1300.4704659852&    1352.1998520434\\
            &$d_{4\x3\x21}$&   0.9220079194& -4.0077609078&     4.1124501332\\
            &$c_{12\x34}$&     -13.7672899406&   -4.3871539915&   14.4494080313\\
            &$c_{4\x123}$&     25.1014609317&   92.8813584055&   96.2134610133\\
            &$c_{3\x412}$&     67.1532044289&  252.1462326711&  260.9353857094\\
 \hline
+\,--\,--\,+&$d_{3\x21\x4}$&    20.8960073185&   19.9656478672&    28.9010417911\\
            &$d_{4\x3\x21}$&   -0.4267971904&   -3.8149302668&   3.8387301002\\
            &$c_{12\x34}$&     -9.0987353955&  -7.6314351405&   11.8754279123\\
            &$c_{4\x123}$&     1.9450686855&   1.9314994054&    2.7411643775\\
 \hline
+\,--\,+\,--&$c_{4\x123}$&      5.2050785566&   1.1864337667&    5.3385829453\\
 \hline
 \end{tabular}
 \caption{Numerical values of coefficients
 of the $\qb q gg\Higgs$ process that differ from ref.~\cite{Budge:2020oyl} at kinematic point, \ref{kinpoint_paper}.}
 \label{aqggtable}
 \end{center}
 \end{table}

 After including the integrals and rational terms, the values of the colour-ordered subamplitudes are
\begin{align}
A^{1234}(1^+,2^+,3^+,4^+;\Higgs)&=-26.505 233 03 - 3.722 078 577 \, i,
 \; & |A^{1234}(1^+,2^+,3^+,4^+;\Higgs)| &= 26.765 299 30 \, , \nonumber \\
A^{1234}(1^+,2^+,3^+,4^-;\Higgs)&=10.005 500 42 + 10.391 302 52 \, i,
 \; & |A^{1234}(1^+,2^+,3^+,4^-;\Higgs)| &= 14.425 297 46 \, , \nonumber \\
A^{1234}(1^+,2^-,3^+,4^-;\Higgs)&=2.105 330 472 - 3.500 785 469 \, i,
 \; & |A^{1234}(1^+,2^-,3^+,4^-;\Higgs)| &= 4.085 084 491 \, , \nonumber \\
A^{1234}(1^+,2^+,3^-,4^-;\Higgs)&=-0.788 758 613 + 0.151 525 137 \, i,
 \; & |A^{1234}(1^+,2^+,3^-,4^-;\Higgs)| &= 0.803 181 185 \, .\nonumber \\
\end{align}
\begin{align}
A^{34}(1^+,2^-,3^+,4^+;\Higgs)&= -3.151 452 974 + 5.766 222 683 \,i, \; 
&|A^{34}(1^+,2^-,3^+,4^+;\Higgs)|=6.571 223 621 \, , \nonumber \\
A^{34}(1^+,2^-,3^-,4^+;\Higgs)&= 1.375 544 184 + 1.088 612 645 \,i, \;
&|A^{34}(1^+,2^-,3^-,4^+;\Higgs)|=1.754 194 771 \, , \nonumber \\
A^{34}(1^+,2^-,3^+,4^-;\Higgs)&= 3.032 201 250 - 1.275 260 855 \,i, \; 
&|A^{34}(1^+,2^-,3^+,4^-;\Higgs)|=3.289 458 111 \, . \nonumber \\
\end{align}
\begin{align}
A^{4q}(1^{+},2^{-},3^{+},4^{-};\Higgs) &= 1.583 011 630 - 1.072 246 795 \,i, \; 
&|A^{4q}(1^{+},2^{-},3^{+},4^{-};\Higgs)|=1.911 972 544 \, . \nonumber \\
\end{align}

\section{Large mass limit}
\label{sec:largemass}

Using Eq.~(\ref{Fermion2gluonresult}) and the large mass expansion for the scalar triangle integral,
\beq
C_0(p_1,p_2;m) \to -\frac{1}{2\, m^2} -\frac{s}{24 m^4} + O\left(\frac{1}{m^6}\right),\;\;\;\;s=\mh^2=2 p_1 \cdot p_2\, ,
\eeq
we can extract the effective interaction for the fermionic theory,
\beq
{\cal L}_{hgg} = -\frac{1}{4} C_f G^{\mu \nu}_a\;G_{\mu \nu\, a} \, h,\;\;\;\; 
C_f =-\frac{g_s^2}{12 \pi^2 v},
\eeq
valid when $\mh^2 \ll m^2$. 
From Eq.~(\ref{Scalar2gluonresult1}) the corresponding effective Lagrangian for the scalar loop is,
\beq
{\cal L}_{hgg} = -\frac{1}{4} C_s G^{\mu \nu}_a\;G_{\mu \nu\, a} \, h,\;\;\;\; 
C_s =\frac{g_s^2}{24 \pi^2 m^2} \Big(-\frac{\lambda}{4}\Big)\,.
\eeq
From these equations we see that,
\beq
\frac{\left(\frac{m^2}{v}\right) C_s}
{\left( \frac{-\lambda}{4} \right) C_f} = -\frac{1}{2} \,.
\label{eq:factors}
\eeq
So, in our canonical normalization in which the coupling factors shown on the left-hand
side of Eq.~(\ref{eq:factors}) are extracted,
the amplitudes for the fermion- and scalar-mediated cases are related
in the large-mass (EFT) limit by a factor of $-1/2$.  In other words,
\bea
m^2 A(\ldots; h) \to -\frac{1}{2} m^2 H(\ldots; h) \,,
\eea
where the asymptotic forms for the fermion-mediated amplitudes
$m^2 H(\ldots; h)$ are given in Appendix B of Ref.~\cite{Budge:2020oyl}.

\bibliography{NotesScalar}
\bibliographystyle{JHEP}

\end{document}